\documentclass[11pt]{article}
\usepackage{amsmath,amssymb,amsfonts,latexsym,epsfig, color, varioref, graphics,setspace,verbatim,bm,amsthm,mathrsfs,multicol,mathtools}
\usepackage{fullpage}
\usepackage{natbib}
\usepackage{booktabs}
\usepackage{bm}
\usepackage{amsthm}
\usepackage{multirow}
\usepackage{caption}
\usepackage{xcolor}
\usepackage{soul}
\DeclarePairedDelimiter{\ceil}{\lceil}{\rceil}
\newcommand\ds{\displaystyle}
\newcommand{\Pn}{ {\mathbb P}_{n}}
\newcommand{\Gn}{ {\mathbb G}_{n}}
\newcommand{\NN}{{\mathbb N}}
\newcommand\ve{\varepsilon}
\newcommand\vp{\varphi}
\newcommand\tg{\textsf{g}}
\newcommand{\wh}{\widehat}

\newcommand{\Bb}{\bm\beta}
\newcommand{\Ba}{\bm \alpha}
\newcommand{\Bg}{\bm\gamma}
\newcommand{\Btau}{\bm\tau}
\newcommand{\Bt}{\bm\theta}
\newcommand{\Bn}{\bm\nu}

\newcommand{\BZ}{\mathbf{Z}}
\newcommand{\BX}{\mathbf{X}}
\newcommand{\BD}{\mathbf{D}}

\newcommand{\BS}{\mathbf{S}}
\newcommand{\BW}{\mathbf{W}}

\newcommand{\ba}{\mathbf{a}}
\newcommand{\bt}{\mathbf{t}}
\newcommand{\bh}{\mathbf{h}}
\newcommand{\p}{\intercal}

\DeclareMathOperator*{\argmax}{\arg\!\max}
\definecolor{brown1}{rgb}{1.00,0.25,0.25}
\definecolor{brown2}{rgb}{0.93,0.23,0.23}
\definecolor{brown3}{rgb}{0.80,0.20,0.20}
\definecolor{brown4}{rgb}{0.55,0.14,0.14}
\definecolor{brown}{rgb}{0.65,0.16,0.16}
\definecolor{green1}{rgb}{0.00,1.00,0.00}
\definecolor{green2}{rgb}{0.00,0.93,0.00}
\definecolor{green3}{rgb}{0.00,0.80,0.00}
\definecolor{green4}{rgb}{0.00,0.65,0.00}
\definecolor{steelblue}{rgb}{0.27,0.51,0.71}
\definecolor{DodgerBlue}{rgb}{0.12,0.56,1.00}
\definecolor{RoyalBlue}{rgb}{0.25,0.41,0.88}
\definecolor{lightgrey}{rgb}{0.5,0.5,0.5}
\definecolor{islamicgreen}{rgb}{0,0.56,0}
\definecolor{bubblegum}{rgb}{0.99, 0.76, 0.8}
\definecolor{lavenderpink}{rgb}{0.98, 0.68, 0.82}
\definecolor{lightsalmonpink}{rgb}{1.0, 0.6, 0.6}
\definecolor{alizarin}{rgb}{0.82, 0.1, 0.26}
\definecolor{carnationpink}{rgb}{1.0, 0.65, 0.79}
\definecolor{coralred}{rgb}{1.0, 0.25, 0.25}
\definecolor{pinksherbet}{rgb}{1, 0.46, 0.55}
\definecolor{blue(ryb)}{rgb}{0.01, 0.28, 1.0}
\definecolor{ao(english)}{rgb}{0.0, 0.5, 0.0}
\definecolor{bleudefrance}{rgb}{0.19, 0.55, 0.91}
\newtheorem{theorem}{Theorem}
\newtheorem{lemma}{Lemma}
\theoremstyle{remark}
\newtheorem{remark}{Remark}

\begin{document}
\setstretch{1.2}
\setlength\parindent{0pt}

\title{An efficient penalized estimation approach for a semi-parametric linear transformation model with interval-censored data}
\author{Minggen Lu, Yan Liu, Chin-Shang Li, Jianguo Sun}
\maketitle

{\bf Abstract}\\
We consider efficient estimation of flexible transformation models with interval-censored data. To reduce the dimension of semi-parametric models, the unknown monotone transformation function is approximated via monotone splines. A penalization technique is used to provide more computationally efficient estimation of all parameters. To accomplish model fitting, a computationally efficient nested iterative expectation-maximization (EM) based algorithm is developed for estimation, and an easily implemented variance-covariance approach is proposed for inference on regression parameters. Theoretically, we show that the estimator of the transformation function achieves the optimal rate of convergence and the estimators of regression parameters are asymptotically normal and efficient. The penalized procedure is assessed through extensive numerical experiments and further illustrated via two real applications.

\textit{Key words}: efficient estimation; interval-censored data; monotone spline; penalized estimation; transformation model

\section{Introduction}
The aim of this study is to develop and evaluate a novel methodology to conduct regression analysis for interval-censored data. Interval-censored data arise naturally in many epidemiological and
biomedical research in which the exact time to an event of interest cannot be measured directly, but rather is known to lie within an interval. For example, in a retrospective breast cancer study (Beadle et al., 1984), the rate of deterioration of the cosmetic result was compared between patients with primary radiation therapy plus adjuvant chemotherapy and patients with radiotherapy alone. Patients were scheduled every 4 to 6 months to inspect the cosmetic state at a clinic visit. The information on time until one of endpoints, the appearance of breast retraction that is highly associated with overall cosmetic deterioration, was only known between two adjacent clinic visits.

The regression analysis of interval-censored data has been extensively investigated in literature. In particular, a lot of work has been focusing on the proportional hazards (PH) model, for example, Finkelstein (1986), Satten (1996), Pan (1999), Betensky et al. (2002), Zhang et al. (2010),  and Wang et al. (2016), among many others. It is noted that the restrictive assumption of proportionality of hazards made in the above work is not always justifiable.   To provide a more general and flexible modeling framework, a lot of effort has been involved in the development of estimation methodology under semi-parametric linear transformation models that include PH model as a special case, and can be classified as an estimating equation method (Zhang et al., 2005; Zhang and Zhao, 2013), fully semiparametric maximum likelihood approach (Zeng et al., 2016), and monotone spline technique (Zhou et al., 2017 and 2018); for a comprehensive review of analysis of interval-censored data see Sun (2006), Zhang and Sun (2010), and the references therein.

The aforementioned methods under transformation models can be either computationally intensive and unstable or complicated to implement. Thus in this work we strive to develop a flexible and computationally efficient penalized spline estimation approach that can be used to fit semi-parametric transformation models with interval-censored data. In particular, we made use of  monotone $B$-spline to model the unknown transformation function and utilized penalization methodology to facilitate the spline estimation.  To accomplish the model fitting, a nested iterative EM based algorithm was developed for estimation, and a simple and consistent variance-covariance estimation approach was proposed to provide accurate and reliable inference for regression coefficients. As demonstrated via extensive numerical studies, the proposed penalized approach is  computationally efficient and robust to model misspecification under a variety of settings.  Theoretically, the large-sample properties such as the optimal rate of convergence of the functional estimator and the asymptotic normality and efficiency of regression parameter estimators were established. These appealing properties both numerically and theoretically emphasize the fact that the proposed penalized procedure contributes to the existing methodology for interval-censored data.

The remainder of this paper is presented as follows. The methodological details of the penalized approach including  the development of a nested iterative EM based algorithm and the selection of the smoothing parameter and spline knots are presented in Section~\ref{estimation}.  The large-sample properties of penalized estimators and a consistent variance-covariance estimation approach  are provided in Section~\ref{asymptotic}. The numerical performance of the penalized method is assessed via the extensive Monte Carlo studies and compared with the existing approach (Wang et al. 2016) under the PH model in Section~\ref{simulation}. The proposed method is further illustrated by analyzing data from a retrospective breast cancer study and a prospective signal tandmobiel study in Section~\ref{realdata}. The summary of the study and future work are discussed  in Section~\ref{discussion}.  	Finally,  the proofs of the asymptotic results  and more simulation results are available  in Supplementary Materials.

\section{Model and numerical algorithm}\label{estimation}
\subsection{Model and penalized likelihood}
Consider a linear transformation model in which the conditional distribution of the true failure time $T$ given the covariate vector $\BZ$ takes the form
\begin{equation}\label{model}
\tg\{F(t|\BZ)\}=\vp(t)+\BZ^\intercal\Bb,
\end{equation}
where $\Bb$ is a $d$-dimensional vector of regression coefficients corresponding to the covariate vector $\BZ$. In this specification, it is assumed that the link function $\tg(\cdot)$ is  a known and increasing function on $(0,1)$ and   the \textit{unknown} transformation function $\vp(\cdot)$ is  smooth and monotone increasing  on $(0,\infty)$ with $\vp(0) = -\infty$. The above formation provides immensely flexible modeling of censored data and includes a group of commonly used models as special cases. For instance, model~\eqref{model} reduces to the proportional hazards (PH) model for specifying $\tg(x) = \log\left\{-\log(1-x)\right\}$, while taking $\tg(x) = \log\left\{x/(1-x)\right\}$ leads to the proportional odds (PO) model.   

In this work we are interested in fitting  model~\eqref{model} to interval-censored data. The fundamental feature of interval-censored data is that the exact event time $T$  cannot be ascertained, but rather is censored by two observation times, say $L$ and $R$ with $L<R$. Let $\Delta_{i1} = I(T<L)$, $\Delta_{i2} = I(L<T<R)$, and $\Delta_{i3} = I(T > R)$ denote the censoring indicators, subject to $\Delta_{1}+\Delta_{2}+\Delta_{3} = 1$, where $I(\cdot)$ is the indicator function. The event time $T$ is left-, interval-, or right-censored if $\Delta_1 = 1$, $\Delta_2=1 $, or $\Delta_3 = 1$, respectively. Consequently, the observed data comprise $\{\Delta_{i1},\Delta_{i2},\Delta_{i3}, L_i,R_i, \BZ_i\}_{i=1}^n$. We assume that  the covariates are time-invariant and that the event time and the censoring process are independent given the covariates. Under these assumptions, the log-likelihood of the observed data for $\Btau = (\Bb,\vp)$ is given by 
\begin{equation}\label{lik}
\ell_n(\Btau)  = \sum_{i=1}^n\left[
\Delta_{i1}\log\{F(R_i|\BZ_i)\}+\Delta_{i2}\log\left\{F(R_i|\BZ_i)-F(L_i|\BZ_i)\right\}+\Delta_{i3}\log\{1-F(L_i|\BZ_i)\}\right].
\end{equation}
Penalized estimation is well-known for its feature to control for the smoothness of nonparametric estimator and to avoid overfitting the model; see Ma and Kosorok (2005) and Lu and Li (2017) and the references therein. Proceeding in this manner, we make use of the following penalized log-likelihood to estimate the parameters in model~\eqref{model} 
\begin{equation}\label{plik}
\ell_{n,\lambda}=\ell_n(\Btau) - \frac{1}{2}\lambda^2J^2(\vp),
\end{equation}
where $J^2(\vp) = \int\left\{\vp^{(r)}(t)\right\}^2dt$ for $r>0$ is the penalized term  and $\lambda>0$  is the smoothing parameter that is used to control for the smoothness of fitted function. 

In general, it is an extremely challenging task to estimate the infinite-dimensional nonparametric function directly from the penalized likelihood.  Following the work of Lu and Li (2017),  to tackle the difficulty, 
we propose to approximate the unknown transformation function $\vp(\cdot)$ via monotone $B$-splines. In particular,
$$
\vp(\cdot) \approx \vp_n(\cdot) =  \sum_{j=1}^{q_n}\gamma_jB_j(\cdot)
$$
under the constraints $\gamma_1\le\ldots\le\gamma_{q_n}$, where $B_j(\cdot), j = 1,\ldots,q_n$, are $B$-spline basis functions. The nondecreasing coefficients $\Bg = (\gamma_1,\ldots,\gamma_{q_n})^\intercal$ guarantee that the resulting spline function $\vp_n(\cdot)$ is nondecreasing; see Theorem 5.9 of Schumaker (1981).  Thus, via spline approximation, we obtain the spline model
\begin{equation}\label{smodel}
\tg\{F_n(t|\BZ)\}= \vp_n(t)+\BZ^\intercal\Bb. 
\end{equation}
Under the above spline model~\eqref{smodel}, the set of unknown parameters that have to be estimated is given by $\Bt=(\Bb,\Bg)$. Moreover, to further relieve the computational burden, we approximate the penalized term $J^2(\vp)$ by the discrete difference penalty; see Eilers and  Marx (1996).  By introduction of discrete penalty, the estimation can be easily fit via standard software. 
Under these specifications, the {\it observed} penalized spline log-likelihood for $\Bt$ is given by
\begin{align}\label{pslik}
l_{o,\lambda}(\Bt) =&\sum_{i=1}^n\left[
\Delta_{i1}\log\{F_n(R_i|\BZ_i)\}+\Delta_{i2}\log\left\{F_n(R_i|\BZ_i)-F_n(L_i|\BZ_i)\right\}+\Delta_{i3}\log\{1-F_n(L_i|\BZ_i)\}\right] \nonumber \\
&-\frac{1}{2}\lambda^2\|\BD\Bg\|^2,
\end{align}
where  {$\|\cdot\|$ is the usual Euclidean norm} and $\BD$ is the matrix representation of the discrete difference operator of order 2;  see Wood (2017, p.168). 

\subsection{The EM algorithm}\label{algorithm}
To accomplish model fitting, motivated by Wang et al. (2016), we developed a reliable and efficient EM based optimization procedure to identify the penalized estimator of $\Bt$.  In particular,  define $\NN_i(t)$ as the latent Poisson process for subject $i$ with mean rate function $H_i(t) = -\log\left[1-\tg^{-1}\left\{\vp_n(t)+\BZ_i^\p\Bb\right\}\right]$, $i=1,\ldots,n$.  Let $T_i$ denote the time to the occurrence of the first event of the Poisson process. Thus,  $T_i$ indeed follows the spline transformation model~\eqref{smodel} with a cumulative distribution function $F_n(t|\BZ_i) = \tg^{-1}\left\{\varphi_n(t)+\BZ_i^\p\Bb\right\}$. 
To implement the EM algorithm for the proposed model, two latent variables need to be specified. In particular, let $Y_i =\NN_i(U_{i1})$, where $U_{i1} = \Delta_{i1}R_i+(1-\Delta_{i1})L_i$, and similarly, when $\Delta_{i1} = 0$, define $W_i = \NN(U_{i2})-\NN(U_{i1})$, where $U_{i2}=\Delta_{i2}R_i+\Delta_{i3}L_i$.  Under these specifications, if $T_i$ is left-censored, then $Y_i>0$, and in  the case of interval-censoring or right-censoring, $Y_i = 0$ and $W_i > 0$ or $Y_i = 0$ and $W_i = 0$, respectively. Observe that $Y_i\sim$Poisson$(\lambda_{i1})$, and  in the case of $\Delta_{i1} = 0$, $W_i\sim$Poisson$(\lambda_{i2})$, where $\lambda_{i1}=H_i(U_{i1})$ and $\lambda_{i2}=H_i(U_{i2})-H_i(U_{i1})$; see Wang et al. (2016) for further information. 
Thus, under these specifications, ignoring the items without including $\Bt$, the complete penalized spline log-likelihood for $\Bt$ involving the latent $Y_i$'s and $W_i$'s as well as observed data is given by
\begin{eqnarray*}\label{clik}
	{l}_{c,\lambda}(\Bt) = \sum_{i=1}^n \left\{Y_i\log\lambda_{i1} -\lambda_{i1}+(\Delta_{i2}+\Delta_{i3})(W_i\log\lambda_{i2} -\lambda_{i2})\right\}-\frac{1}{2}\lambda^2\|\BD\Bg\|^2.
\end{eqnarray*}
During the E-step of the EM algorithm, conditional on the observed data $\mathcal{O}_i$, $i=1,\ldots,n$, and the current estimate $\Bt^{(m)} = (\Bb^{(m)},\Bg^{(m)})$ of $\Bt$, taking the expectation of $l_{c,\lambda}(\Bt)$ with respect to all the latent variables $Y_i$'s and $W_i$'s yields the penalized $Q$ function, namely,
$$
Q_\lambda(\Bt,\Bt^{(m)}) = \sum_{i=1}^n \left[ E^{(m)}(Y_i)\log\lambda_{i1} -\lambda_{i1}+(\Delta_{i2}+\Delta_{i3})\left\{E^{(m)}(W_i)\log\lambda_{i2} -\lambda_{i2}\right\}\right] -\frac{1}{2}\lambda^2\|\BD\Bg\|^2,
$$
where $E^{(m)}(Y_i) \equiv E(Y_i|\mathcal{O}_i;\Bt^{(m)})$ and $E^{(m)}(W_i) \equiv E(W_i|\mathcal{O}_i;\Bt^{(m)})$. 
Under the specification of $Y_i$, $Y_i>0$ if and only if $\Delta_{i1} =1$, and, given $\mathcal{O}_i$ and $Y_i>0$, $Y_i$ is a zero-truncated Poisson distribution. Thus, 
$$
E(Y_i|\mathcal{O}_i;\Bt^{(m)}) =  \frac{H_i^{(m)}(R_i)}{1-\exp\{-H_i^{(m)}(R_i)\}}\Delta_{i1},
$$
where $H_i^{(m)}(t)$ represents $H_i(t)$ evaluated at $\Bt = \Bt^{(m)}$. 
Similarly, under the specification of $W_i$, we have
$$
E(W_i|\mathcal{O}_i;\Bt^{(m)}) =  \frac{H_i^{(m)}(R_i) - H_i^{(m)}(L_i)}{1-\exp\left\{-H_i^{(m)}(R_i)+H_i^{(m)}(L_i)\right\}}\Delta_{i2}.
$$       
The M-step involves finding $\Bt^{(m+1)}= \argmax_{\Bt}Q_\lambda(\Bt,\Bt^{(m)})$ under the constraints $\gamma_1\le\ldots\le\gamma_{q_n}$ for a fixed $\lambda$. The optimization can be easily accomplished via applying the standard linearly constrained optimization routines such as \textsf{R} function \textsf{constrOptim} that is based on an adaptive barrier algorithm. 
\begin{remark}
By taking advantage of $B$-spline approximation, the proposed EM algorithm simplifies the alternative of Wang et al. (2016) for the PH and PO models and extends it to the  transformation models.  
\end{remark}
 
\subsection{Nested optimization procedure}
In what follows, we summarize a nested iterative optimization procedure that involves in updating estimates of both the parameters and the smoothing parameter. The outer process is committed to streamlining the smoothing  parameter $\lambda$, while the inner process is aimed at solving a constrained penalized  function for a fixed value $\lambda$, which can be achieved via applying an efficient EM algorithm as discussed in Section~\ref{algorithm}. 
At this point, we identify the maximizer of the penalized spline log-likelihood function  $\wh\Bt_\lambda^{(m)}$ for a fixed $\lambda$. 
Once the inner process is finished,  we implement the generalized Fellner-Schall method  (Wood and Fasiello, 2017) to update the estimate of the smoothing parameter in the outer process.  In particular, the updated estimate of $\lambda$, say $\overline{\lambda}$, is given by
\begin{equation}\label{update}
\overline{\lambda}^2 =
\frac{\text{tr}\{\BS^-_\lambda\BS\}-\text{tr}\left\{\mathcal{J}_{n,\lambda}^-(\wh\Bt_\lambda^{(m)})\BS\right\}}{\wh\Bt^{(m)\intercal}_\lambda\BS\wh\Bt_\lambda^{(m)}}\lambda^2,
\end{equation}
where $\BS^-_\lambda$ is the generalized inverse of $\BS_\lambda = \lambda\BS$ with  $\BS = \mbox{diag}\{\mathbf{0}_{d\times d},\BD^\p\BD\}$  and $\mathcal{J}_{n,\lambda}^{-}(\wh\Bt_\lambda^{(m)})$ is the generalized inverse of $\mathcal{J}_{n,\lambda}(\wh\Bt_\lambda^{(m)})$, the negative Hessian matrix of the observed penalized spline log-likelihood function~\eqref{pslik} evaluated at $\Bt_\lambda = \wh\Bt_\lambda^{(m)}$. Because $\mathcal{J}_{n,\lambda}(\Bt_\lambda)$ is non-negative definite, the numerator in~\eqref{update} is guaranteed to be non-negative; see Theorem 4 of Wood and Fasiolo (2017) for further information. 
Essentially,  formula~\eqref{update}  is used to update the estimate of the smoothing parameter each time after $\wh\Bt_\lambda^{(m)}$ is available. Once the updated estimate of the smoothing parameter, i.e., $\overline{\lambda}$, is available, the inner process is then restarted to identify $\wh{\Bt}^{(m)}_{\overline{\lambda}}$. These two processes iterate in turn until  convergence, i.e., the absolute difference between $\Bt_\lambda^{(m)}$ and $\Bt_{\overline{\lambda}}^{(m)}$ is less than a pre-specified number. The implementation of the nested iterative algorithm is outlined as follows: 
\vspace{-8mm}
\begin{description}
\item[Step 1 (outer process).] Identify $\wh{\Bt}_\lambda^{(m)}$ for the current $\lambda$ through the EM algorithm discussed in Section \ref{algorithm}. 
\item[Step 2 (inner process).] Update $\lambda$ to $\overline{\lambda}$ via equation~\eqref{update} and go back to Step 1 to identify $\wh\Bt_{\overline{\lambda}}^{(m)}$.  Two processes iterate until $\|\wh\Bt_{\overline{\lambda}}^{(m)}-\wh\Bt_\lambda^{(m)}\| < 10^{-6}$.
\end{description}\vspace{-4mm}
When the algorithm converges, $\wh{\Bt} = (\wh{\Bb},\wh{\Bg})$ is defined as the penalized spline  estimator of $\Bt$. Accordingly, $\wh\vp(\cdot)= \vp_n(\cdot|\wh\Bg)$ is the penalized spline estimator of $\vp(\cdot)$. 

It is noted that once the smoothing parameter is specified,  the regression parameters and the spline coefficients can be estimated simultaneously using the existing procedure.  Moreover, as demonstrated via the extensive simulation studies in Section~\ref{simulation}, the proposed algorithm usually converges in a few steps and is robust to the selection of the initial values and spline knots. Further, the issue of divergence is very rare.  These findings reinforce the assertion that the penalized spline estimation approach is computationally  stable and efficient. 

\subsection{Selection of knot}
A sensitivity analysis displayed in Table 1 reveals that the proposed penalized procedure is robust to the selection of knots. Thus, we choose the number of inner knots to be $\left\lceil n^{1/3}\right\rceil$ denoting the least integer greater than or equal to $n^{1/3}$. Once the number of knots is fixed, the location of knots is determined by equally-spaced quantiles of the observation times.  This empirical rule was applied in the simulation study and the real application. 

\section{Asymptotic results}\label{asymptotic}
Let $\Btau_0 = (\Bb_0,\vp_0)$  and $\wh\Btau = (\wh\Bb,\wh\vp)$ denote the true value and the penalized estimator of $\Btau$, respectively.  Define the $L_2$-norm of $\vp$ as
$$
\|\vp_2-\vp_1\|_2^2 = E\{\vp_2(L) - \vp_1(L)\}^2+E\{\vp_2(R) -
\vp_1(R)\}^2.
$$
We establish the asymptotic properties of $\wh\Btau$ with $L_2$-norm
$$
\|\Btau_2 - \Btau_1\|_2^2 = \|\Bb_2 - \Bb_1\|^2 + \|\vp_2 - \vp_1\|_2^2. 
$$
Write the log-likelihood of $\BX = (L,R,\BZ,\Delta_1,\Delta_2,\Delta_3)$ for $\Btau$ as
$$
\ell(\Btau) =\Delta_1\log\pi_1+\Delta_2\log(\pi_2 - \pi_1) +\Delta_3\log(1-\pi_2),
$$
where $\pi_1 = \tg^{-1}(\zeta_1)$ and $\pi_2 = \tg^{-1}(\zeta_2)$ with $\zeta_1 = \vp(L)+\BZ^\intercal\Bb$ and $\zeta_2 = \vp(R)+\BZ^\intercal\Bb$. The score function for $\Bb$  takes the form
$$
\dot\ell_{\Bb}(\Btau) =\frac{D(\zeta_1)\BZ}{\pi_1}\Delta_1+\frac{D(\zeta_2)\BZ-D(\zeta_1)\BZ}{\pi_2-\pi_1}\Delta_2-\frac{D(\zeta_2)\BZ}{1-\pi_2}\Delta_3,
$$
where $D(\zeta) = d\tg^{-1}(\zeta)/d\zeta$. Differentiating $\ell(\Bb,\vp_{\bt})$ along a parametric smooth model $\vp_{\bt} = \vp+\bt^\p \bh$ at $\bt = 0$ for $\bh=(h_1,\ldots,h_d)^\intercal$ with $h_l\in\mathcal{H} = \left\{h: E\{h^2(\cdot)\}<\infty, J(h)<\infty\right\}$, $1\le l\le d$, we have score operator for $\vp$
$$
\dot\ell_\vp(\Btau)[\bh] =\frac{D(\zeta_1)\bh(L)}{\pi_1}\Delta_1+\frac{D(\zeta_2)\bh(R)-D(\zeta_1)\bh(L)}{\pi_2-\pi_1}\Delta_2-\frac{D(\zeta_2)\bh(R)}{1-\pi_2}\Delta_3.
$$	
Let $\bh^*$ denote the least favorable direction that minimizes the distance $\|\dot{\ell}_{\Bb}(\Btau) - \dot\ell_\vp(\Btau)[\bh] \|_2^2$ for $\bh\in\mathcal{H}^d$, i.e., $\dot\ell_\vp(\Btau)[\bh^*]$ is the orthogonal projection of $\dot\ell_{\Bb}(\Btau)$ onto the space $\mathcal{H}^d$.  
The information matrix for $\Bb$ at $\Btau = \Btau_0$ is then defined as
$$
\mathscr{I}(\Bb_0)=E\{\ell^*_{\Bb}(\Btau_0) \}^{\otimes 2}\equiv E\{ \dot{\ell}_{\Bb}(\Btau_0)  - \dot{\ell}_\vp(\Btau_0)[\bh^*]\}^{\otimes 2},
$$
where $\ell^*_{\Bb}(\Btau_0)$ is the efficient score for $\Bb$ at $\Btau = \Btau_0$. The following regularity conditions are sufficient to establish the asymptotic results.\vspace{-0.35cm}
\begin{enumerate}
\item[(A1)] The true regression parameter vector $\Bb_0$ is an interior point of a compact subset of $\mathbb{R}^d$.
\item[(A2)] The true transformation function $\vp_0(\cdot)$ is strictly increasing with $0<1/M < \vp_0(d_1)<\vp_0(d_2)<M$ for some $M>0$, where $0<d_1<d_2<\infty$ and its $r$th derivative satisfies  Lipschitz condition on $[d_1,d_2]$ with $r\ge 1$.
\item[(A3)] The first derivative of link function $\tg(\cdot)$ is bounded away from 0 and infinite and the second derivative of $\tg(\cdot)$ is bounded on $(0,1)$. 
\item[(A4)] (a) There exists a positive number $\eta$ such that $P(R - L\ge\eta) = 1$, and (b) the union of the supports of $L$ and $R$ is contained in $[d_1,d_2]$.
\item[(A5)] (a) The covariate vector $\BZ$ has a bounded support,  and (b) $\Pr(\BZ^\intercal\Bb\neq\BZ^\intercal\Bb_0)>0$ for any $\Bb\neq\Bb_0$.
\item[(A6)] The joint density of $(\BZ,L,R)$ is bounded away from 0 and infinite.  
\item[(A7)] The smoothing parameter is assumed to be of order
$$
\lambda= O_p(n^{-r/(1+2r)})~~~\mbox{and}~~~\lambda^{-1} = O_p(n^{r/(2r+1)}).
$$
\item[(A8)] The information matrix $\mathscr{I}(\Bb_0)$ is positive definite.
\end{enumerate}

\begin{theorem} \label{rate}(Consistency and rate of convergence)\label{Convergence} Under conditions A1 -- A7, the penalized estimator $\wh\Bb$ is consistent for $\Bb_0$, $\|\wh\vp\|_\infty =O_p(1)$, and $\|\wh\vp-\vp_0\|_2=O_p(n^{-r/(1+2r)})$.
\end{theorem}

\begin{theorem} (Asymptotic normality and efficiency)\label{Normality} Under conditions A1--A8,
$$
n^{1/2}(\wh\Bb-\Bb_0) 
\xrightarrow{\enskip d \enskip}\mathcal{N}(0,\mathscr{I}^{-1}(\Bb_0)), ~~n\rightarrow\infty. 
$$
\end{theorem}
\begin{remark}
Theorem~\ref{Convergence} suggests that for a choice of $\lambda$ that is of the order
$O_p(n^{-r/(1+2r)})$, $\wh\vp(\cdot)$ attains the optimal rate of convergence $O_p(n^{r/(1+2r)})$ in the semi-parametric setting. Theorem \ref{Normality} states that the penalized  estimator $\wh\Bb$ achieves the information bound and, hence, is efficient; see Bickel et al. (1993) for more discussion of semiparametric efficiency.   
\end{remark}
As discussed in Huang and Wellner (1997) and Zhang et al. (2010), the least favorable direction $\bh^*$ is the solution of a Fredholm integral equation of the second kind and does not have a closed form. Therefore,  it is very challenging to estimate $\bh^*$ and, hence, the information matrix $\mathscr{I}(\Bb_0)$ directly. Following the proposal of Zhang et al. (2010), by taking advantage of the orthogonal property of the least favorable direction and $B$-spline approximation, we put forward a least-square based approach to tackle the problem.  In particular, in view of definition of $\bh^*$, define $\wh\bh^* = \min_{\bh\in\mathcal{H}_n^d}\|\dot\ell_{\Bb}(\wh\Btau) - \dot\ell_{\vp}(\wh\Btau)[\bh]\|_n^2$,  where $\|\cdot\|_n$ is the usual empirical norm and $\mathcal{H}_n$ is an approximation $B$-spline space for $\mathcal{H}$. 
Following the arguments in Zhang et al. (2010), it can be shown that $\wh\bh^*$ is well-defined and can be used to consistently estimate $\bh^*$. It follows that
$$
\wh{\mathcal{I}}_n \equiv\Pn\{\dot{\ell}_{\Bb}(\wh{\Btau}) -\dot{\ell}_{\vp}(\wh{\Btau})[\wh{\bh}^*]\}^{\otimes 2} 
$$
is a consistent estimator of $\mathscr{I}(\Bb_0)$, where $\Pn$ is the empirical measure. It is worthwhile to point out that $\wh{\mathcal{I}}_n$  is the outer product version of the observed information for $\Bb$ without penalized term. Let  $\wh{\mathcal{I}}_{n,\lambda}$ denote the outer product version of the {\it penalized} observed information for $\Bb$. The following theorem results from the consistency of $\wh{\mathcal{I}}_n$ and the assumption of the smoothing parameter. 
\begin{theorem} (Variance estimation)
\label{var} Under conditions C1--C8, $\wh{\mathcal{I}}_{n,\lambda}$ is an asymptotically consistent estimator of ${\mathscr{I}}(\Bb_0)$.
\end{theorem}\vspace{-4mm}
The finite-sample performance of the proposed variance-covariance estimation method was evaluated in the Monte Carlo study and employed for the inference on regression parameters in the real application.

\section{Simulation}\label{simulation}
In this section we assess the proposed methodology under a variety of simulation settings and compare it with an existing approach (Wang et al. 2016) implemented in $\mathsf{R}$  package $\mathsf{ICsurv}$. 
Assume the true failure time was generated from a class of transformation models
$$
\tg_{\alpha}\{F(t|\BZ) \} =\vp(t)+Z_1\beta_1 + Z_2\beta_2,
$$
where $Z_1\sim$ Bernoulli(0.5) and $Z_2\sim \mathcal{N}(0,1)$. To demonstrate the broad applicability of the proposed approach, we assume that the link function belongs to a class of smooth and monotone increasing functions indexed by a parameter $\alpha$ as follows:
\begin{align}\label{transfun}
\tg_\alpha(u) =
\begin{cases}
\ds\log\frac{(1-u)^{-\alpha}-1}{\alpha}       & \quad \text{if}\ \alpha > 0, \\
\log\{-\log(1-u)\}  & \quad \text{if}\ \alpha=0.
\end{cases}
\end{align}
It is noted that the family of  models presented above include the most commonly used survival models, i.e., $\alpha=0$ or $\alpha=1$ corresponding to the PH or PO model, respectively. To evaluate the performance of the proposed methodology across a wide range of settings, we considered three simulation configurations that differ in the regression parameters $\Bb$ and the unknown transformation function $\vp(\cdot)$. In particular, in configuration I (C1),  $\vp(t)=\log\{(t^2+t)/5 \}$ and $\Bb=(-1,-1)^\intercal$; in configuration II (C2), $\vp(t)=\log(t)$ and $\Bb=(-1,1)^\intercal$; and in configuration III (C3), $\vp(t)=\log\{\log(1+3t)+t/3\}$ and $\Bb=(1,-1)^\intercal$. Under each configuration, three different models were considered,  i.e., $\alpha=0$ (PH), $\alpha=0.5$, and $\alpha=1$ (PO).  To generate interval-censored data, the number of observations for each individual was simulated from 1 plus a Poisson random variable with expectation 1 for the purposes of avoiding zero observation, and the gap times between two consecutive observations follow an exponential distribution with mean parameter 0.5. These specifications provide a  diversity of right-censored rates, i.e., for $\alpha = 0, 0.5$, and 1, the right-censored rates are 74\%, 76\%, and 78\%, respectively, under C1, 41\%, 46\%, and 51\%, respectively, under C2, and  14\%, 21\%, and 27\%, respectively, under C3.  For each considered configuration, 1,000 datasets were generated under each model with sample size $n=50$ or $n=100$.  Cubic $B$-splines with the knots being equally-spaced percentiles were employed to model the unknown transformation function $\vp(\cdot)$. 

To demonstrate robust performance of the penalized method to the specification of the number of knots, a sensitive analysis was conducted with the number of inner knots being selected to be 3, 5, and 7. Table 1 presents a summary of simulation results for the regression parameters (i.e., $\beta_1$ and $\beta_2$) under C1, across the three considered transformation models (i.e., $\alpha=0,0.5$, and 1) and two different sample sizes (i.e., $n = 50$ or 100). The summaries of simulation results for C2 and C3 are provided in the Web Tables 1a and 1b, respectively. One would observe that, under the three considered simulation settings, the estimation and inference for $\Bb$ across all considered numbers of knots are almost the same, suggesting the robustness of the penalized spline method to  the selection of the number of knots. Therefore, it is  suitable in our simulation study and real application to choose the number of inner knots to be $n^{1/3}$.  
These results reveal that the penalized approach exhibits desirable numerical properties 
across all considered settings. In particular,  the empirical biases are small and approach to 0 as the sample size increases, providing numerical assertion of asymptotic consistency. Moreover,  the standard deviations (SD) of the point estimates are in accordance with the averages of the estimated standard errors (ASE) derived from  the proposed variance-covariance estimation method, and the corresponding empirical coverage probabilities of 95\% Wald-type confidence intervals are close to  the nominal level. These findings give an impression that the large-sample inference based on the proposed variance-covariance estimation procedure is feasible under all considered models, even with a relatively small sample size. At last, the power of the Wald-type test for the regression parameters based on the proposed variance-covariance estimation procedure was examined. On the basis of the change of power for $\Bb$ under {C1} of the PO model displayed in Figure 1, it suggests that the proposed procedure can be employed to accurately estimate the power of Wald-type test, i.e., the estimated powers display the symmetry around true values and increase when the sample size or effect size is increased.

We compared the proposed methodology with the competing $I$-spline based sieve semi-parametric approach outlined in Wang et al. (2016) under the PH model. There are many reasons to choose this approach for a direct comparison. In particular, this competing method is the most recent contribution to the existing methodology for interval-censored data and was implemented in $\mathsf{R}$ package $\mathsf{ICsurv}$. For purposes of comparison, cubic splines are used for both the penalized spline and the $I$-spline methods. For a particular data set, as discussed above, the penalized approach selects the number of knots as a cubic root of sample size. In contrast, the competing method makes use of varying configurations of the number of knots. The $I$-spline based sieve estimator is identified via some model selection criteria such as the Akaike's information criterion (AIC); see Wang et al. (2016) for more details. Table 2 summarizes the result of comparison for estimations of the regression parameters under C1. The results of comparisons under C2 and C3 are provided in Web Tables 2a and 2b, respectively.  From this summary, the proposed penalized estimation method exhibits smaller mean squared errors and better empirical coverage probabilities when compared with the competing approach. Further, the standard error estimates obtained from the proposed variance-covariance estimation procedure display less variability when compared to the ones based on the Louis's method implemented in R package \textsf{ICsurv}. It was found that the competing $I$-spline approach suffers from the issue of algorithm divergence and the negative variance estimate, especially when the number of knots is large or the right censoring rate is high. Therefore, it is evident that the proposed methodology outperforms the approach of Wang et al. (2016) under the PH model across all considered settings even through the competing sieve estimator was selected as the best from multiple knot configurations. 

The results with respect to the estimation of the transformation function $\vp(\cdot)$ under C1 of the PH (i.e., $\alpha = 0$) and the PO (i.e., $\alpha=1$) models for all considered sample sizes are summarized in Figure 2. Included in the figure are pointwise means and the corresponding 2.5\% and 97.5\% percentiles of the 1,000 estimates, together with the true value of the function. Web Figures 2a and 2b exhibit similar results under C2 and C3, respectively.  It is observed that the spline curve fits the true function very well and the pointwise lower and upper limits of 95\% confidence interval are fairly close, indicating negligible bias and little variability of the spline estimates. Further, both bias and variability decrease with sample size, which provides a numerical verification to the asymptotic consistency established in Section~\ref{asymptotic}. 

Finally, we investigated the robustness of the proposed methodology in the case where the assumption of PH or PO is not valid, i.e., select the PH model $(\alpha = 0)$ for $\alpha = 0.2$ and PO model ($\alpha = 1$) for $\alpha = 0.8$ under C1.  As displayed in Table 3, the proposed methodology still exhibits the satisfactory finite-sample performance in terms of bias, standard deviation, average of estimated standard errors, and coverage probability, i.e., the approach is robust to the specification of the link function to a certain extent.  The analogous results under C2 and C3 are available in Web Tables 3a and 3b, respectively. In summary, the proposed penalized method displays desirable numerical properties such as accurate estimation and reliable inference even for a small sample size under a variety of simulation settings and  models for interval-censored data.

\section{Real data application}\label{realdata}
\subsection{Breast cosmesis study}
For purposes of illustration, we applied the proposed methodology to a breast cosmesis study (Beadle et al., 1984).  The aim of this retrospective study was to assess the effect of adjuvant chemotherapy on cosmetic state for early breast cancer patients who received primary radiation therapy. 
Two treatment regimens were assigned, 
i.e., 48 patients were assigned to the radiation therapy plus adjuvant chemotherapy, and 46 to the radiation therapy alone. Each patient was scheduled to visit the clinic every 4 to 6 months to check overall cosmetic appearance. For this analysis, the event of interest is the appearance of breast retraction which is highly correlated to overall cosmetic deterioration. It is noted that the exact  time until the appearance of breast retraction cannot be observed attribute to the design of the study.  It was only known that the event time occurred within an interval, i.e., it is interval-censored.  Further,  as discussed in Finkelstein and Wolfe (1985), the time between clinic visits was unlikely to be related to the development of cosmetic deterioration, i.e., the non-informative censorship  assumption was appropriate for analysis of this data set.  

For this analysis, we were interested in comparing the rate of cosmetic deterioration between two treatments. 
Cubic $B$-spline with $\ceil{n^{1/3}} = 5$ inner knots placed at equally-spaced percentiles of follow-up time was utilized to model the transformation function. 
The nested algorithm outlined in Section~\ref{algorithm} was employed for model fitting, and 
the proposed variance-covariance estimation method discussed in Section~\ref{asymptotic} was applied to estimate the standard errors of the regression parameter estimators. Table 4 summarizes the result including the regression parameter estimates and the estimated standard errors along with the corresponding 95\% confidence intervals under the PH and the PO models. Both models suggest that adjuvant chemotherapy is a significant risk factor for cosmetic deterioration. In particular, patients who received chemotherapy in addition to primary radiation treatment demonstrated a higher hazard of developing breast retraction when compared to those who received primary radiation therapy alone under the PH model. The similar result is observed for the PO model, i.e., the chemotherapy increases the odds of breast retraction for those who have previously received the primary radiation therapy. These results support the findings provided in Finkelstein and Wolfe (1985), Finkelstein (1986), and Zhang et al. (2010). The estimated transformation functions along with 95\% pointwise confidence intervals under the PH and the PO models are displayed in Figure 3. The lower and upper limits of 95\% pointwise confidence intervals were calculated as 2.5\% and 97.5\% percentiles of 1,000 pointwise bootstrap estimates, respectively. 

\subsection{Signal tandmobiel study}
To further illustrate the proposed approach, we analyzed data from signal tandmobiel study that is a longitudinal prospective dental study conducted in Northern Belgium from 1996 to 2001. In this study, a total of 4,468 randomly selected children (2,315 boys and 2,153 girls) who were born in 1989 were examined annually (at most 6 times) by one of the 16 trained dentists. 
At each clinic visit, the information on the emergence of permanent tooth (i.e., ``emerged'' versus ``not emerged'') was recorded. In addition, the health status of each erupted tooth was inspected and made into two categories: ``sound'' versus ``caries experience'' (i.e., presence of caries, filled, or extracted for reasons of caries). Because the participants were examined on an annual basis, caries experience data were interval-censored. It is noted that clinical visits were scheduled independently of the time of caries experience, i.e., it is reasonable to assume the non-informative censoring for this study. For this analysis, we aimed to detect the association between caries experience and three potential risk factors, i.e., gender, type of education system (free school, community school, or province/council school), and the starting age of brushing the teeth. 
In particular, we are interested in the 4 first molar teeth, whose right-censoring rates are around 25\%. In our analysis, we analyzed the health status of one of the maxillary first molar teeth (i.e., tooth 26); see p. 29 of Bogaerts et al. (2017) for more information of tooth identification number) by using the following model
\begin{equation*}
\tg\{F(t|Z) \} =\vp(t)+\beta_1 Z_1+\beta_2Z_2+\beta_3Z_3+\beta_4Z_4,
\end{equation*}
where $\tg(\cdot)$ is a known link function.  
In the above specifications,  $Z_1 =1$ for boy and $Z_1 = 0$ for girl, $Z_2=1$ if the child attended community school ($Z_2=0$ otherwise), $Z_3=1$ if the child attended province/council school ($Z_3=0$ otherwise), and $Z_4$ indicates the starting age of brushing the teeth (as reported by parents), i.e., the nearest integer that is greater than the true age. 
The unknown transformation function $\vp(\cdot)$ was approximated by cubic $B$-splines. The selection of knots and smoothing parameter follows the discussion outlined in Section~\ref{algorithm}.  Model fitting was accomplished through the nested algorithm summarized in Section~\ref{algorithm},  and the inference on the regression parameters was based on the proposed variance-covariance approach summarized in Section~\ref{asymptotic}. Presented  in Table 5 are regression parameter estimates and estimated standard errors along with corresponding 95\% confidence intervals under the PH and the PO models.  Both models suggest that starting to brush the teeth at a later age is a significant risk factor for caries experience, while gender and school attendance are not. In particular,  participants who started brushing the teeth at a later age experienced a significant earlier caries for an erupted tooth. Further, the estimates of the transformation function under the PH and the PO models along with 95\% pointwise confidence intervals are shown in Figure 4. The lower and upper limits of the 95\% pointwise confidence intervals are determined by 2.5\% and 97.5\% percentiles of 1,000 bootstrap estimates.

\section{Summary and discussion}
\label{discussion}
In this work a computationally efficient penalized approach has been developed to analyze interval-censored data under flexible semi-parametric linear transformation models. The proposed nested iterative algorithm is straightforward to implement via existing software. The  proposed penalized estimation method exhibits desirable numerical properties that are validated through the extensive Monte Carlo studies. Further, a  simple and consistent variance-covariance estimation approach was developed to provide reliable inference for regression parameters. In addition to the appealing numerical properties, the nonparametric estimator achieves the optimal rate of convergence, and the regression parameter estimators are shown to be asymptotically normal and efficient.  


Inspired by the attainment of the proposed penalized approach, future work will be targeted at extending the proposed methodology to a partially linear additive transformation model for interval-censored data and a linear transformation model for heavily right-censored interval-censored data. 
Another desirable direction for future research could involve the development of an approach to identifying the link function. One option is to apply the single-index methodology to transformation models, i.e., treat the monotone link function as unknown and model it via penalized monotone splines. Proceeding in this way allows us to fully capture the nonparametric feature of the link function. Due to the complex nature of interval-censored data and single-index model this is an extremely challenging topic. A lot of effort would be involved in the discussion of the identifiability of the model and the development of numerical algorithm  and large-sample properties of the penalized estimators.




\section{Appendix}
\subsection{Notations}Let $P$ and $P_0$ be the distribution of $\BX$ for $\Btau$ and $\Btau_0$, respectively, with respect to a $\sigma$-finite measure $\mu$, and $p_{\Btau}$ and $p_0$ be the corresponding densities. The empirical process evaluated at a measurable function $f$ is defined as $\Gn{f}\equiv\sqrt{n}(\Pn-P)f$, where $\Pn{f}$ is the expectation of $f$ under the empirical measure $\Pn$.  
Accordingly, define $\|\mathbb{G}_n\|_{\mathcal{F}}\equiv\sup_{f\in\mathscr{F}}|\mathbb{G}_nf|$ for a class of measurable functions $\mathscr{F}$.  Let $N(\delta,\mathscr{F},L_q(P)$ and $N_{[]}(\delta,\mathscr{F},L_q(P))$ denote the covering number and the bracketing number with $\mathscr{F}$ for $L_q(P)$-metric with $q\ge 1$ ($L_q(P)$ reduces to the supremum norm $\|\cdot\|_\infty$ for $q=\infty$). In the sequel, the expressions $a\lesssim b$ and  $a\gtrsim$ b represent $a\le b$ and $a\ge b$, respectively, up to a constant. We apply Theorem 25.81 of van der Vaart (2000) to derive the consistency and the rate of convergence of $\wh\Btau$. Define the criterion function in above theorem as
$$
m_{\Btau,\lambda} = m_{\Btau} - \frac{1}{2}\lambda^2 \left\{J^2(\vp) - J^2(\vp_0)\right\},
$$
where $m_{\Btau} = \log\{(p_{\Btau}+p_{0})/(2p_{0})\}$.

\subsection{Technical lemmas}
\begin{lemma}\label{inequality}
If conditions A1--A7 hold, then  
$$
\int(p_{\Btau}-p_0)^2d\mu\gtrsim \|\Btau-\Btau_0\|_2^2,
$$
for $\Btau$ in a neighborhood of $\Btau_0$. 
\end{lemma}

\begin{lemma}\label{inf}
If conditions A1--A7 hold, then, for a sufficiently small
$\eta> 0$,  
$\|\vp\|_\infty\lesssim 1+J(\vp)$  whenever $\|\vp-\vp_0\|_2 < \eta$.
\end{lemma}

\begin{lemma}\label{entropy}
If conditions A1--A7 hold, then, for $\eta>0$, 
$$
\sup_P\log N\left(\ve,\left\{m_{\Btau}:
\Bb\in\mathbb{R}^d, J(\vp)\le \eta\right\},L_2(P)\right)\lesssim \left(\dfrac{1+\eta}{\ve}\right)^{1/r}.
$$
\end{lemma}

\begin{lemma}\label{normality}
Suppose that the following conditions are assumed\vspace{-0.3cm}
\begin{enumerate} 
\item[(B1)] $\Pn\ell_{\Bb}^*(\wh\Btau) = o_p(n^{-1/2})$,
\item[(B2)] $(\Pn-P)\{\ell_{\Bb}^*(\wh\Btau) - \ell_{\Bb}^*(\Btau_0)\} = o_p(n^{-1/2})$,
\item[(B3)] $P\{\ell_{\Bb}^*(\wh\Btau) - \ell_{\Bb}^*(\Btau_0)\} = -\mathscr{I}(\Bb_0)(\wh\Bb-\Bb_0)+o_p(n^{-1/2})$. 
\end{enumerate}\vspace{-0.3cm}
Then
$$
\sqrt{n}(\wh\Bb-\Bb_0) = \sqrt{n}\mathscr{I}^{-1}(\Bb_0)\Pn\ell^*_{\Bb}(\Btau_0)+o_p(1)\xrightarrow{\enskip d \enskip} N(0,\mathscr{I}^{-1}(\Bb_0)),~~n\rightarrow\infty.
$$
\end{lemma}

\subsection{Proof of Lemma~\ref{inequality}} 
Let $\ell_{\Btau}=\log p_{\Btau}$ and $\ell_0 = \log p_{\Btau_0}$. The density function $p_0$ is bounded away from a sufficiently small constant $\eta>0$ and infinite 
and, hence, $p_{\Btau}$ is also bounded away form a positive constant and infinite, for $\Btau$ varying in a small neighborhood of $\Btau_0$. It follows that $\ell_{\Btau}$ and $\ell_0$ are bounded.  Then, by the mean value theorem, $\int(p_{\Btau}-p_0)^2d\mu$ is bounded below by $E(\ell_{\Btau}- \ell_0)^2$, up to a constant, and, hence, $\int(p_{\Btau}-p_0)^2d\mu\gtrsim E(\ell_0 - \ell_{\Btau})$. Let $\pi_{10}$ and $\pi_{20}$ denote $\pi_1$ and $\pi_2$ evaluated at $\Btau = \Btau_0$, respectively. By some straightforward algebra, we have
\begin{align*}
P_0(\ell_0-\ell_{\Btau}) & = E\left\{\pi_{10}\log\frac{\pi_{10}}{\pi_1}+(\pi_{20}-\pi_{10})\log\frac{\pi_{20}-\pi_{10}}{\pi_2-\pi_1}+(1-\pi_{20})\log\frac{1-\pi_{20}}{1-\pi_2}\right\}\\
&=E\left\{\pi_1m\left(\frac{\pi_{10}}{\pi_1}\right) +(\pi_2-\pi_1)m\left(\frac{\pi_{20} - \pi_{10}}{\pi_2 - \pi_1}\right) +(1-\pi_2)m\left(\frac{1-\pi_{20}}{1-\pi_2}\right) \right\},
\end{align*} 
where $m(x) = x\log x-x+1\ge 1/4(x-1)^2$ for $x$ in a neighborhood of 1. It follows that  
\begin{align*}
E(\ell_0-\ell_{\Btau}) & \gtrsim E\left\{\frac{1}{\pi_1}(\pi_1-\pi_{10})^2+ \frac{1}{1-\pi_2} (\pi_2-\pi_{20})^2\right\}\\
&\gtrsim E\left[\left\{\xi_1(\BZ)+\xi_2(L)\right\}^2+ \left\{\xi_1(\BZ)+\xi_2(R)\right\}^2\right],
\end{align*}
where $\xi_1(\BZ)=\BZ^\intercal(\Bb-\Bb_0)$ and $\xi_2(t) = \vp(t) - \vp_0(t)$. The last inequality holds because the first derivative of the link function $\tg(\cdot)$ is bounded away from 0 and infinite. Invoke the law of total expectation and Cauchy-Schwarz inequality and then apply the orthogonal property of conditional expectation to conclude that
there exists $0<C_0<1$ such that $\left\{E\xi_1(\BZ)\xi_2(L)\right\}^2 \le C_0 E\xi_1^2(\BZ)E\xi_2^2(L)$. It concludes from Lemma 25.86 of van der Vaart (2000) that
$$
E\{\xi_1(\BZ)+\xi_2(L)\}^2\gtrsim
E\xi_1^2(\BZ)+E\xi_2^2(L).
$$
The above inequality also holds for
$E\{\xi_1(\BZ)+\xi(R))\}^2$ by replacing $L$ by $R$.
Lemma 1 follows.

\subsection{Proof of Lemma~\ref{inf}} 
\noindent  For any $J(\vp)<\infty$, by a Taylor expansion, for $t\in[d_1,d_2]$,
\begin{eqnarray*}
\vp(t) &=& \vp(d_1)+\vp^\prime(d_1)t+\cdots+\frac{\vp^{(r-1)}(d_1)}{(r-1)!}t^{r-1}+\int_{d_1}^t\frac{\vp^{(r)}(u)(t-u)^{r-1}}{(r-1)!}du\\
&\equiv& p(t)+h(t), 
\end{eqnarray*}
where $p(\cdot)$ is a $(r-1)$th degree polynomial and $\sup_{t\in[d_1,d_2]} |h(t)|\le C_0 J(\vp)$ for $C_0>0$ by the Cauchy-Schwarz inequality.  To simplify the presentation, assume $0<C_0<1$, i.e., $\|h\|_\infty\le J(\vp)$. Write $p(t) = \ba^\p\bt$ for $\bt = (1,t,\ldots,t^{r-1})^\p$ and $\ba = (a_0,a_1\ldots,a_{r-1})^\p$ with $a_l = \vp^{(l)}(d_1)/l!$, for $0\le l\le r-1$. Under the assumption $\|\vp - \vp_0\|_2\le\eta$, we have $\|\vp \|_2\le \|\vp_0\|_2+\|\vp - \vp_0\|_2\le M+\eta$ for $M>0$. Then
$$
\frac{\|p\|_2}{1+J(\vp)} \le \frac{\|\vp\|_2}{1+J(\vp)}+\frac{\|h\|_2}{1+J(\vp)} \lesssim M+\eta+1.
$$
The non-singularity of matrix $E(\bt\bt^\p)$ implies that $\|\ba\|_2/\{1+J(\vp)\} \le O(1)$ and, hence, $\|p\|_\infty/\{1+J(\vp)\} \le O(1)$ because $p(\cdot)$ is defined on a bounded set. The Lemma follows from the inequality $\|\vp\|_\infty\le \|p\|_\infty+\|h\|_\infty$.   

\subsection{Proof of Lemma \ref{entropy}}
As shown in Lemma \ref{inf}, for any $J(\vp)<\infty$, $\vp(\cdot)$ can be written as $p(\cdot)+h(\cdot)$, where $p(\cdot)$ is a polynomial of degree $r-1$ and $h$ satisfies $\|h\|_\infty \le J(\vp)$ and $J(h)=J(\vp)$. Let $\mathscr{G}$ be a class of functions with $\|h\|_\infty\le \eta$ and $J(h)\le \eta$. According to Lemma 2.4 of van der Geer (2000), the entropy number $\log N(\ve,\mathscr{G},\|\cdot\|_\infty)$ is of the order $(\eta/\ve)^{1/r}$. For a fixed function $h(\cdot)$,  because $\tg^{-1}(\cdot)$ is one-to-one,  by Example 3.7.4d of van der Geer (2000) and Lemma 9.7 of Kosorok (2008),  the class of bounded functions $\tg^{-1}\{\BZ^\p\Bb+p(t)+h(t)\}$ is a Vapnik-Chervonenkis subgraph class of indexes bounded by $d+r+2$. Lemma 19.15 of van der Vaart (2000) applies and concludes that the uniform covering number of the above class for $L_2$-norm is of order $(1/\ve)^{2(d+r+1)}$. Because $J(\vp)\le\eta$ implies $\|h\|_\infty\le \eta$ and $J(h)\le\eta$, it follows that the uniform covering number of the class of functions $\tg^{-1}\{\BZ^\p\Bb+\vp(t)\}$ with $J(\vp)\le \eta$ is of order $(1/\ve)^{2(r+d+1)}\exp\{(\eta/\ve)^{1/r}\}$ and, hence,
$$
\sup_P\log N(\ve,\{\tg^{-1}\{\BZ^\p\Bb+\vp(t)\}: \Bb\in\mathbb{R}^d, J(\vp)\le \eta\},L_2(P))\lesssim \left(\dfrac{1+\eta}{\ve}\right)^{1/r}.
$$
Because $p_{\Btau}$ is essentially a transformation of $\tg^{-1}(\zeta_1)$ and $\tg^{-1}(\zeta_2)$ for $\zeta_1 = \BZ^\p\Bb+\vp(L)$ and $\zeta_2 = \BZ^\p\Bb+\vp(R)$ along with binary censoring indicators $\Delta$, and  $m_{\Btau}$ is Lipschitz with respect to $(\zeta_1,\zeta_2)$, Lemma 3 follows by applying the similar arguments in the proof of  Lemma 9.13 of Kosorok (2008).  

\subsection{Proof of Lemma \ref{normality}}
Combing (B1)--(B3) yields
$$
\Pn \ell_{\Bb}^*(\Btau_0) = \mathscr{I}(\Bb_0)(\wh\Bb-\Bb_0)+o_p(n^{-1/2}).
$$
The lemma follows from the non-singularity of $\mathscr{I}(\Bb_0)$ and  the standard central limit theorem. 	

\subsection{Proof of rate of convergence} 
\noindent  
Because the logarithmic function is concave,  by definition of $\widehat\Btau$, we have
$
\Pn m_{\wh\Btau,\lambda}\ge 0 = \Pn m_{\Btau_0,\lambda}.
$
Following the same arguments in the proof of Lemma 1.3 of van der Geer (2000) yields
$$
\int \log\frac{p_{\Btau}+p_{0}}{2p_{0}}p_{0}d\mu \le  -\frac{1}{2} \int\left( \sqrt{p_{\Btau}+p_0}-\sqrt{2p_{0}}\right)^2d\mu. 
$$
Further, $p_{0}$ and $p_{\Btau}$ are bounded below by positive numbers; see proof of Lemma~\ref{inequality}. It follows that the right-handed term in the above inequality can be bounded by $-\int(p_{\Btau}-p_0)^2d\mu$, up to a constant. We conclude from Lemma~\ref{inequality} that 
$$
P_0(m_{\Btau,\lambda} - m_{\Btau_0,\lambda}) \lesssim  -\|\Btau-\Btau_0\|_2^2 -
\lambda^2J^2(\vp) + \lambda^2.
$$
Hence, condition (25.79) of Theorem 25.81 of van der Vaart (2000) holds with $d^2_{\lambda}(\Btau,\Btau_0) = \|\Btau -
\Btau_0\|_2^2+\lambda^2J^2(\vp)$. According to the assumption of $\lambda$, $d_\lambda(\Btau,\Btau_0)\le\eta$ implies $\|\Bb-\Bb_0\|\le\eta$ and $J(\vp)\le \eta/\lambda_n$ for $\lambda_n = n^{-r/(1+2r)}$. Define $\mathscr{Fz}_\eta = \{m_{\Btau}, \|\Bb-\Bb_0\|\le\eta, J(\vp)\le \eta/\lambda_n\}$. In view of 
Lemma~\ref{entropy}, the uniform entropy integral  for $\mathscr{F}_\eta$ with a given envelop function $F$ is given by
$$
J(1,\mathscr{F}_\eta,L_2) = \int_0^1\sqrt{\log\sup_P N(\ve\|F\|_2,\mathscr{F}_\eta,L_2(P))}d\ve\lesssim \left(1+\frac{\eta}{\lambda_n}\right)^{1/(2r)}\equiv \phi_n(\eta).
$$
We conclude from Theorem 2.14.1 of van der Vaart and Wellner (1996) that condition (25.80) of Theorem 25.81 of van der Vaart (2000) holds. Clearly, $\phi_n(\eta)/\eta^{1/(2r)}$ is decreasing in $\eta$ and $\phi_n(\eta_n)\lesssim\sqrt{n}\eta_n^2$ for $\eta_n = (n^r\lambda_n)^{-1/(4r-1)}$.  It follows from Theorem 25.81 of van der Vaart (2000) that
$d_{\lambda}(\widehat\Btau,\Btau_0)=O_p(\eta_n+\lambda_n)$, which implies the consistency of $\widehat\Btau$, and, hence, $\|\wh\vp\|_\infty=O_p(J(\widehat\vp)+1)$ according to Lemma~\ref{inf}.  

In the sequel, in view of the consistency of $\wh\Btau$, we restrict $\Btau$ to a small neighborhood of $\Btau_0$. If $d_{\lambda}(\Btau,\Btau_0)<\eta$ for $\lambda\ge\lambda_n$,
by Lemma 2, $\|\vp\|_\infty\lesssim\eta/\lambda_n+1$.  By the bracketing results for the parametric class and the Sobolev class, see Example 19.7 and Example 19.10 of van der Vaart (2000), respectively,  the entropy with bracketing of the class of functions $\BZ^\p\Bb+\vp(t)$ with $\|\Bb-\Bb_0\|\le \eta$, $J(\vp)\le \eta$, and $\|\vp\|_\infty\le \eta$ for $L_2$-norm is bounded by $\left\{(1+\eta)/\ve\right\}^{1/r}$, up to a constant. Further, $m_{\Btau}$ is Lipschitz with respect to $(\zeta_1,\zeta_2)$. Thus,
$$
\log N_{[]}(\ve,\mathscr{F}_\eta,L_2(P))\lesssim \left(\frac{1+\eta/\lambda_n}{\ve}\right)^{1/r}
$$
and the corresponding bracketing integral
$$
J_{[]}(\eta,\mathscr{F}_\eta,L_2(P)) = \int_0^\eta\sqrt{\log N_{[]}(\ve,\mathscr{F}_\eta,L_2(P))}d\ve\lesssim \left({1+\eta/\lambda_n}\right)^{1/(2r)}\eta^{1-1/(2r)}.
$$
We conclude from Lemma 19.36 of van der Vaart (2000) that condition (25.80) in Theorem 25.81 of van der Vaart (2000) holds with
$$
\phi_n(\eta)=J(\eta,\mathscr{F}_\eta,L_2(P))\left(1+\frac{J(\delta,\mathscr{F}_\eta,L_2(P))}{\eta^2n^{1/2}}\right).
$$
Here we have used the fact that $\Gn m_{\Btau,\lambda} = \Gn m_{\Btau}$. Further, $\phi_n(\eta)/\eta$ is decreasing in $\eta$ and $\phi_n(\eta_n)\lesssim \sqrt{n}\eta_n^2$ for $\eta_n = \lambda_n$.  Therefore, Theorem 25.81 of van der Vaart (2000) applies and yields $\|\widehat\Btau - \Btau_0\|_2= O_p(\lambda_n)$ and, hence, $J(\wh\vp) = O_p(1)$ and $\|\wh\vp\|_\infty = O_p(1)$.

\subsection{Proof of asymptotic normality and efficiency}
Inserting $(\wh\Bb+\Bn,\wh\vp-\Bn^\p\bh^*)$ for $\Bn\in\mathbb{R}^d$ into the penalized  log-likelihood function $\ell_{n,\lambda}(\Btau)$ and differentiating it at $\Bn=0$, we obtain the stationary equations
$$
0 = \Pn\ell^*_{\Bb}(\wh\Btau)-\lambda^2\int\wh\vp(t)\{\bh^*(t)\}^{(r)}dt.
$$
The Cauchy-Schwarz inequality and $J(\wh\vp) = O_p(1)$ along with the assumption of $\lambda$ apply and yield that condition (B1) holds.  As shown in the proof of Theorem~\ref{rate}, the class of functions $\BZ^\p\Bb+\vp(t)$ with $\|\Bb-\Bb_0\|\le \eta$, $J(\vp)\le \eta$, and $\|\vp\|_\infty\le \eta$ for $\eta>0$ is a Donsker class. Further, the efficient function $\ell_{\Bb}^*(\Btau)$ is Lipschitz with respect to $(\zeta_1,\zeta_2)$. We conclude from the Donsker preservation theorem (Theorem 2.10.6 of van der Vaart and Wellner 1996) that the class of functions $\ell^*_{\Bb}(\Btau)$ with $\|\Bb-\Bb_0\|\le \eta$, $J(\vp)\le \eta$, and $\|\vp\|_\infty\le \eta$ for $\eta>0$ is a Donsker class. By Theorem~\ref{rate}, $\ell^*_{\Bb}(\wh\Btau)$ is contained in the above class. Moreover, by the dominated convergence theorem and the consistency of $\wh\Btau$, $\|\ell^*_{\Bb}(\wh\Btau)-\ell^*_{\Bb}(\Btau_0) \|_2\rightarrow 0$. It follows from Theorem 19.24 of van der Vaart (2000) that condition (B2) holds. Finally, in view of the fact that $P\{\ell_{\Bb}^*(\Btau)\dot\ell_{\vp}(\Btau)[h]\} = 0$ for any $h(\cdot)$ with $J(h)<\infty$ and the rate of convergence of $\wh\Btau$, a Taylor expansion applies and yields that
$$
P\{\ell^*_{\Bb}(\wh\Btau)-\ell^*_{\Bb}(\Btau_0)\} = -\mathscr{I}(\Bb_0)(\wh\Bb-\Bb_0)+o_p(n^{-1/2}).
$$
We conclude the asymptotic normality and the efficiency of $\wh\Bb$ from Lemma~\ref{normality}.

\clearpage
\noindent{\bf Table 1}. Summary of results for knot selection study when $\alpha = 0$ (PH model), $\alpha = 0.5$, and $\alpha = 1$ (PO model) under C1. The results include the number of knots ($m_n$), bias, standard deviation (SD), mean
squared error (MSE), average of estimated standard errors (ASE),
and estimated coverage probability (CP95) of 95\% confidence
interval obtained from 1,000 replications with $n=50$ or 100.
\label{table1}\vspace{-0.5cm}
{\small
\begin{flushleft}
\begin{singlespace}
\begin{tabular}{lllrrrrrrrrrr}
\hline
\hline
& & &\multicolumn{5}{c}{$n=50$}& \multicolumn{5}{c}{$n = 100$}\\
\cmidrule(r){4-8}\cmidrule(l){9-13}
Model& &  &Bias & SD & ASE& MSE & CP95&Bias & SD & ASE& MSE & CP95\\
\cmidrule(r){1-13}
$\alpha = 0$ &         
          &${m_n}=3$&-0.152 &0.881 &0.720 &0.799 &94.5\%&-0.071 &0.506 &0.470 &0.261 &94.3\%\\
&$\beta_1$&${m_n}=5$&-0.149 &0.865 &0.713 &0.769 &94.7\%&-0.066 &0.500 &0.466 &0.254 &94.9\%\\
&         &${m_n}=7$&-0.159 &0.826 &0.699 &0.706 &95.1\%&-0.065 &0.492 &0.462 &0.246 &94.9\%\\
\cmidrule(r){2-8}\cmidrule(l){9-13}
&         &${m_n}=3$&-0.171 &0.490 &0.426 &0.269 &96.1\%&-0.083 &0.294 &0.270 &0.093 &94.0\%\\
&$\beta_2$&${m_n}=5$&-0.169 &0.490 &0.426 &0.268 &95.8\%&-0.079 &0.294 &0.269 &0.093 &94.0\%\\
&         &${m_n}=7$&-0.141 &0.490 &0.418 &0.260 &94.2\%&-0.069 &0.295 &0.267 &0.092 &93.4\%\\
\cmidrule(r){1-13}
					
$\alpha = 0.5$&         
          &${m_n}=3$&-0.175 &0.926 &0.808 &0.886 &95.6\%&-0.081 &0.582 &0.539 &0.345 &95.0\%\\
&$\beta_1$&${m_n}=5$&-0.168 &0.914 &0.803 &0.862 &96.1\%&-0.077 &0.581 &0.536 &0.344 &95.1\%\\
&         &${m_n}=7$&-0.178 &0.898 &0.792 &0.837 &96.2\%&-0.084 &0.578 &0.532 &0.340 &94.9\%\\
\cmidrule(r){2-8}\cmidrule(l){9-13}
&         &${m_n}=2$&-0.171 &0.609 &0.468 &0.400 &94.6\%&-0.085 &0.345 &0.305 &0.126 &94.2\%\\
&$\beta_2$&${m_n}=5$&-0.173 &0.616 &0.469 &0.410 &94.1\%&-0.085 &0.346 &0.305 &0.127 &94.2\%\\
&         &${m_n}=8$&-0.148 &0.608 &0.462 &0.391 &93.4\%&-0.074 &0.345 &0.303 &0.125 &93.8\%\\
\cmidrule(r){1-13}
					
$\alpha = 1$ &         
          &${m_n}=3$&-0.168 &0.925 &0.879 &0.884 &96.8\%&-0.065 &0.611 &0.592 &0.377 &95.5\%\\
&$\beta_1$&${m_n}=5$&-0.165 &0.927 &0.876 &0.886 &96.4\%&-0.065 &0.608 &0.590 &0.373 &95.8\%\\
&         &${m_n}=7$&-0.183 &0.928 &0.872 &0.894 &96.7\%&-0.070 &0.610 &0.589 &0.376 &95.7\%\\
\cmidrule(r){2-8}\cmidrule(l){9-13}
&         &${m_n}=3$&-0.172 &0.585 &0.505 &0.372 &95.6\%&-0.076 &0.348 &0.332 &0.127 &96.2\%\\
&$\beta_2$&${m_n}=5$&-0.178 &0.592 &0.506 &0.382 &95.7\%&-0.081 &0.352 &0.333 &0.130 &95.9\%\\
&         &${m_n}=7$&-0.171 &0.601 &0.504 &0.389 &95.2\%&-0.080 &0.353 &0.332 &0.131 &95.5\%\\[0.5em]
\hline\hline
\end{tabular}
\end{singlespace}
\end{flushleft}
}
	
\clearpage
\noindent {\bf Table 2}. Summary of results for comparison between the penalized and competing approaches for the PH model ($\alpha = 0$) under C1.  Results include bias, standard deviation (SD), mean squared error (MSE), average and standard deviation of estimated standard errors (ASE and SDSE), empirical coverage probability (CP95) of 95\% confidence interval obtained from 1,000 replications with $n=50$ or 100. 
\label{table2}
\vspace{-0.35cm}
\begin{center}
\begin{singlespace}
{
\begin{tabular}{llrrrr}
	\hline
	\hline \\[-0.5em]
	& &\multicolumn{2}{c}{Competing}& \multicolumn{2}{c}{Penalized}\\
	\multicolumn{2}{c}{PH ($\alpha=0$)} &\multicolumn{2}{c}{Approach}& \multicolumn{2}{c}{Approach}\\
	\cmidrule(r){3-4}\cmidrule(l){5-6}
	& &$\beta_1$   & $\beta_2$  & $\beta_1$ & $\beta_2$ \\[0.5em]
	$n =50       $   & Bias     &-0.164 &-0.202 &-0.152 &-0.184  \\
	$             $   & SD      &0.845 &0.508 &0.815 &0.498  \\
	$             $   & MSE     &0.741 &0.299 &0.687 &0.281  \\
	$             $   & ASE     &0.772 &0.449 &0.707 &0.425  \\
	$             $   & SDSE    &0.457 &0.210 &0.162 &0.132  \\
	$             $   & CP95    &92.7\% &92.6\% &95.1\% &96.1\%    \\[0.5em]
	\hline \\[-0.5em]
	$n =100  $   & Bias         &-0.073 &-0.084 &-0.066 &-0.059  \\
	$             $   & SD      &0.516 &0.292 &0.501 &0.290  \\
	$             $   & MSE     &0.271 &0.092 &0.255 &0.087  \\
	$             $   & ASE     &0.506 &0.298 &0.458 &0.266  \\
	$             $   & SDSE    &0.173 &0.118 &0.055 &0.047  \\
	$             $   & CP95    &94.2\% &94.6\% &94.2\% &94.5\%    \\[0.5em]
	\hline\hline
\end{tabular}
}
\end{singlespace}
\end{center}
\vspace{-0.5cm}

\newpage
\begin{center}
	\includegraphics[width = 3.1 in, height = 2.8in]{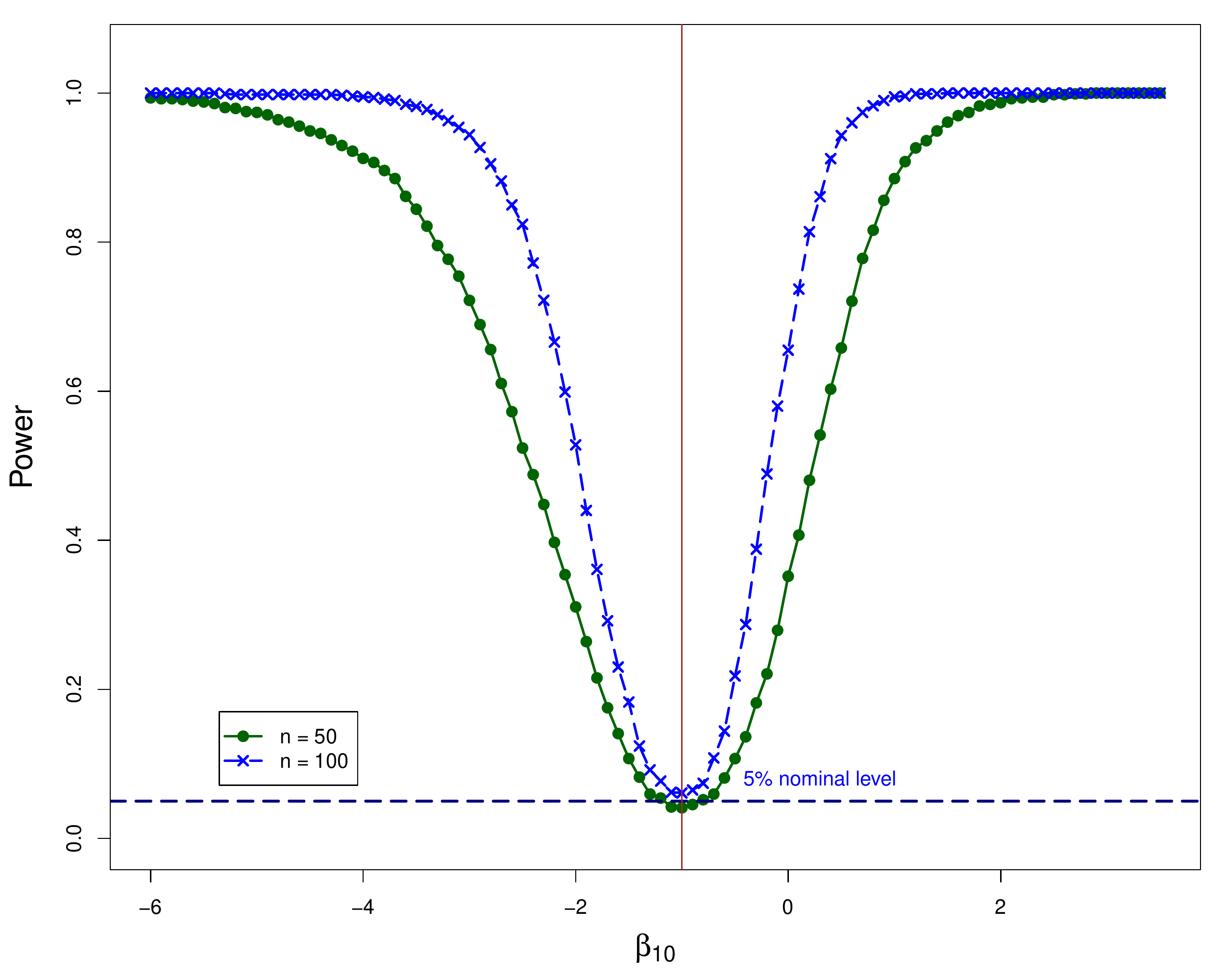}
	\includegraphics[width = 3.1 in, height = 2.8in]{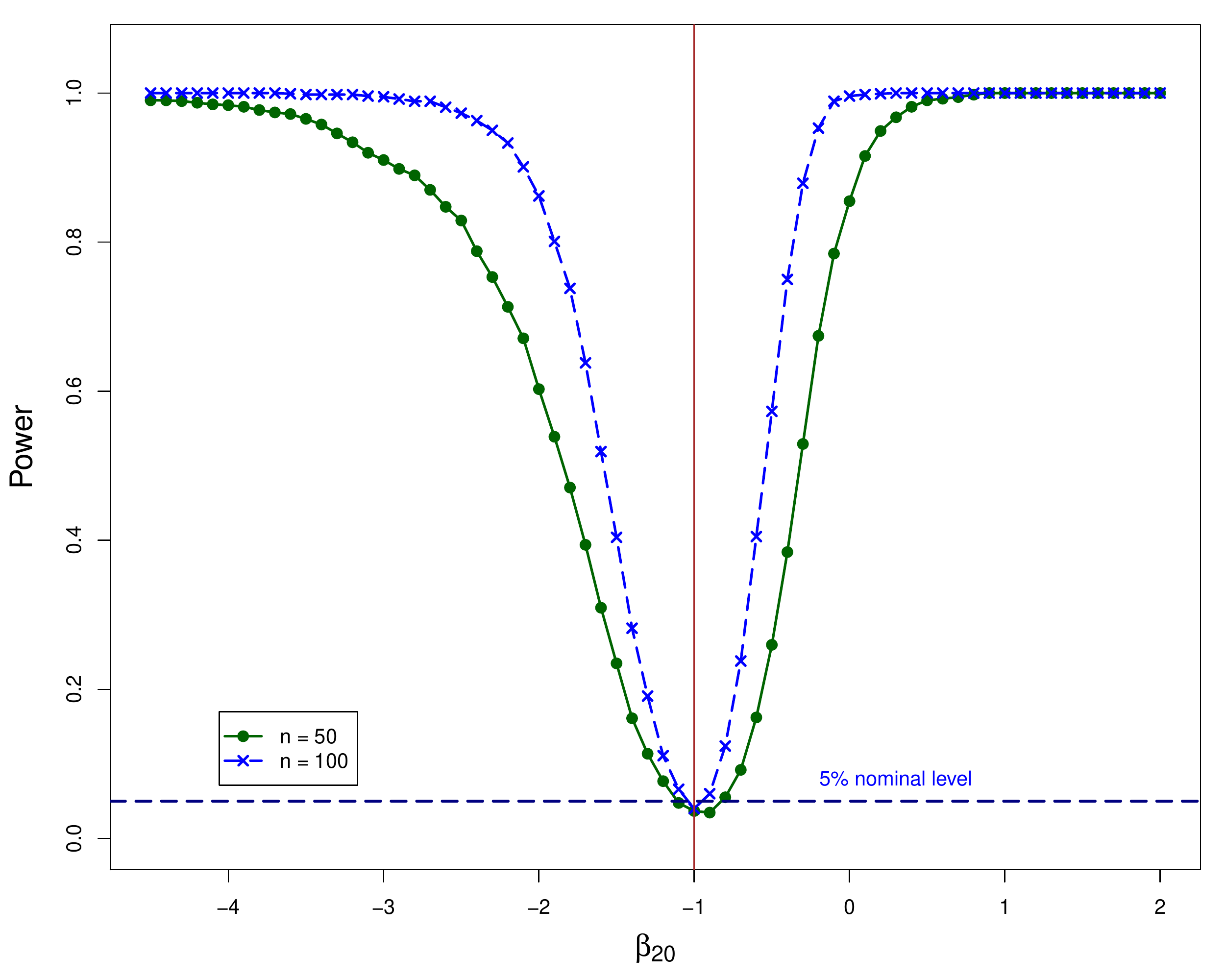}
	\includegraphics[width = 3.1 in, height = 2.8in]{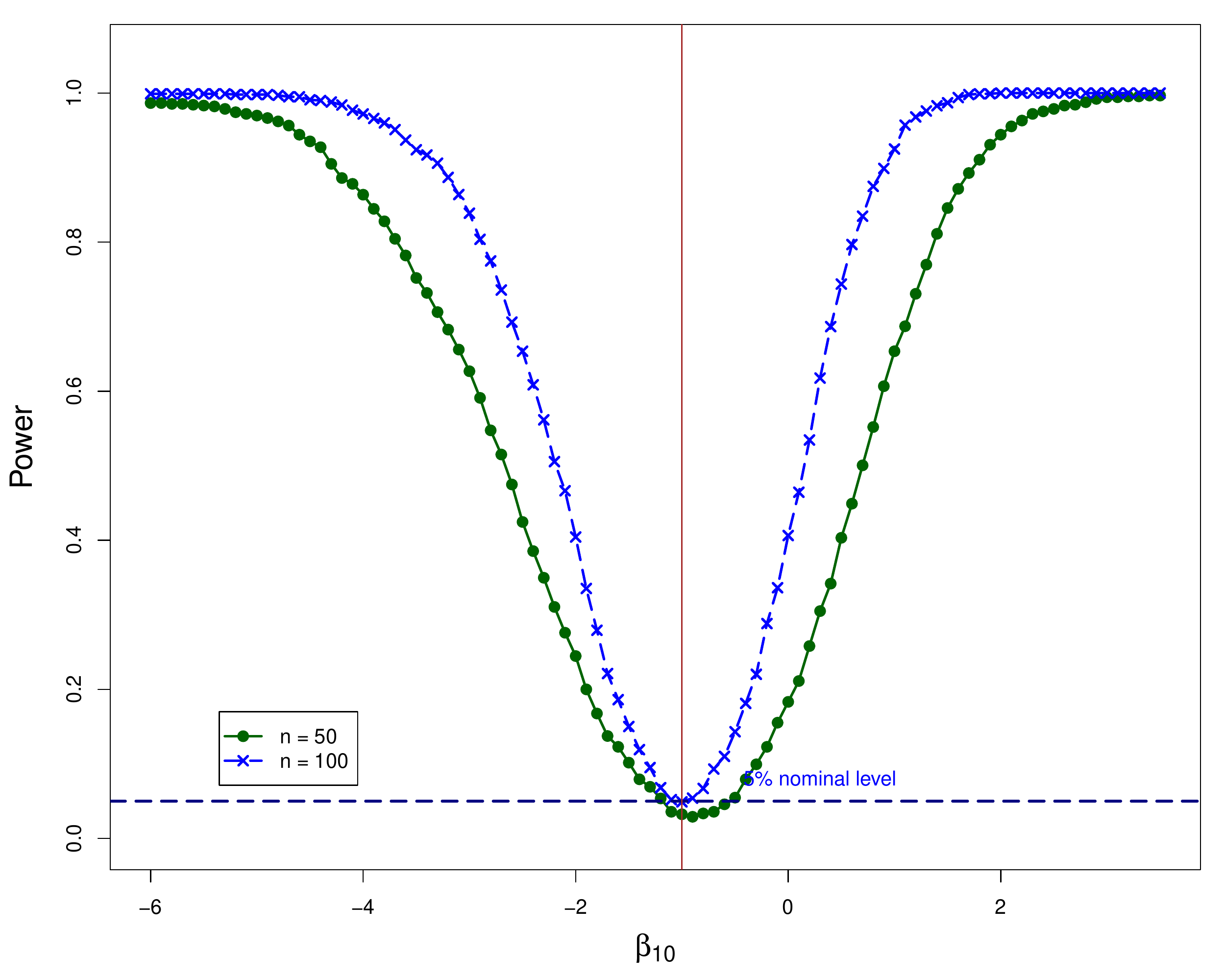}\hspace{1em}
	\includegraphics[width = 3.1 in, height = 2.8in]{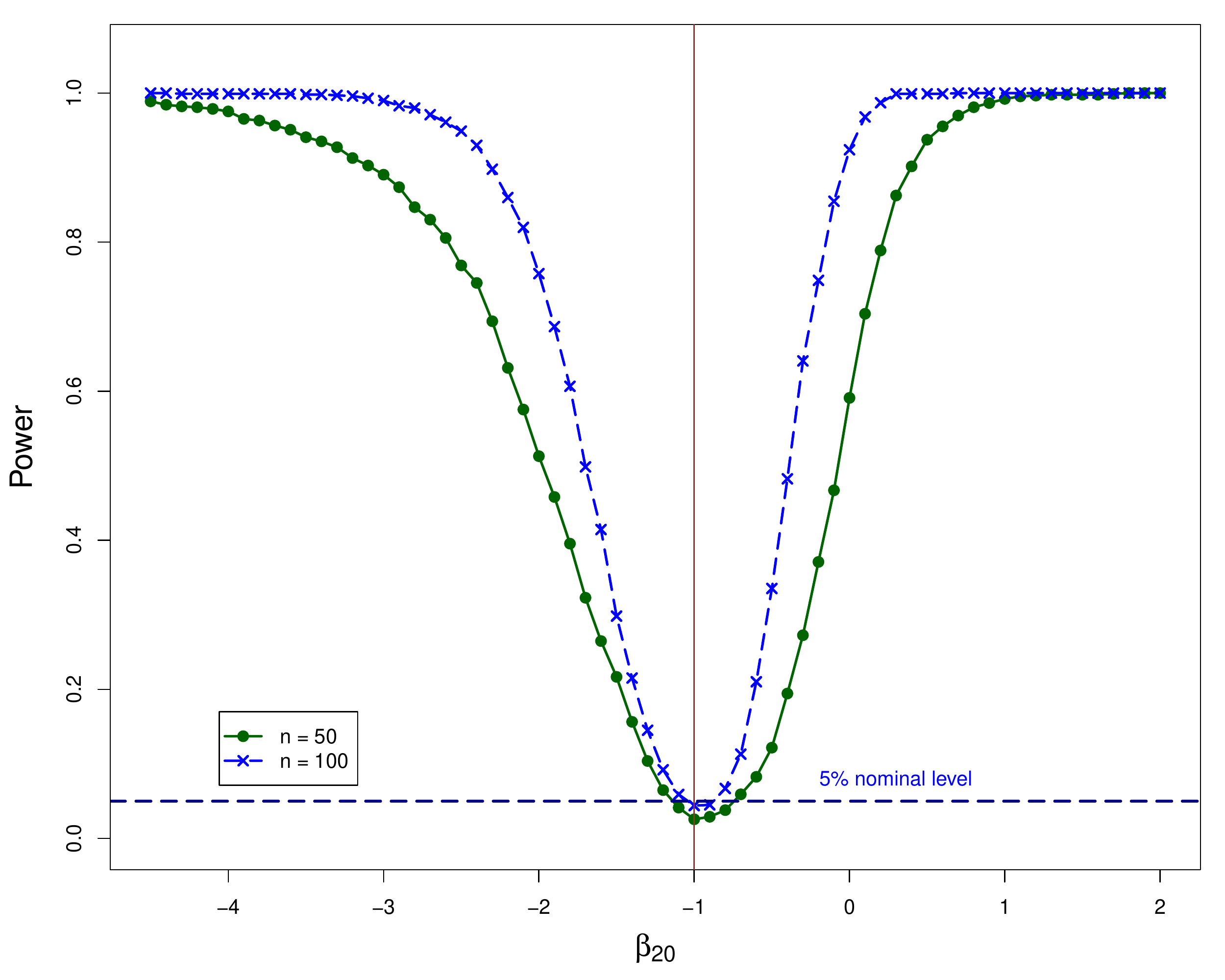}
\end{center}
{\bf Figure 1}. Summary of estimated power curves for $\Bb$ based on 1000 replications for the PH ($\alpha = 0$) (top row) and the PO ($\alpha = 1$) models (bottom row) under C1. These results include estimated powers for $n=50$ (solid curve) and for $n=100$ (long dashed curves).

\newpage
\begin{center}
	\includegraphics[width = 3.1 in, height = 2.8in]{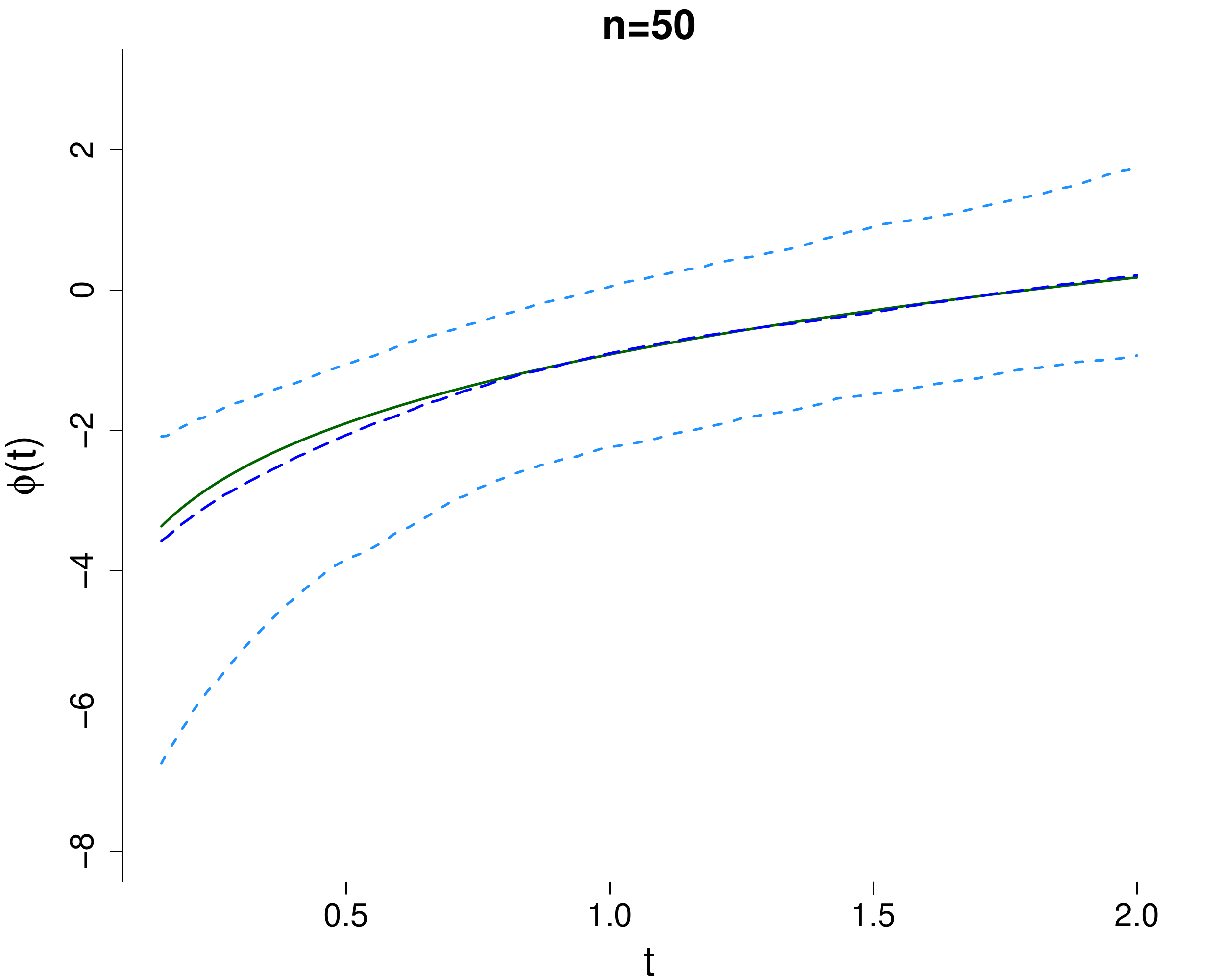}
	\includegraphics[width = 3.1 in, height = 2.8in]{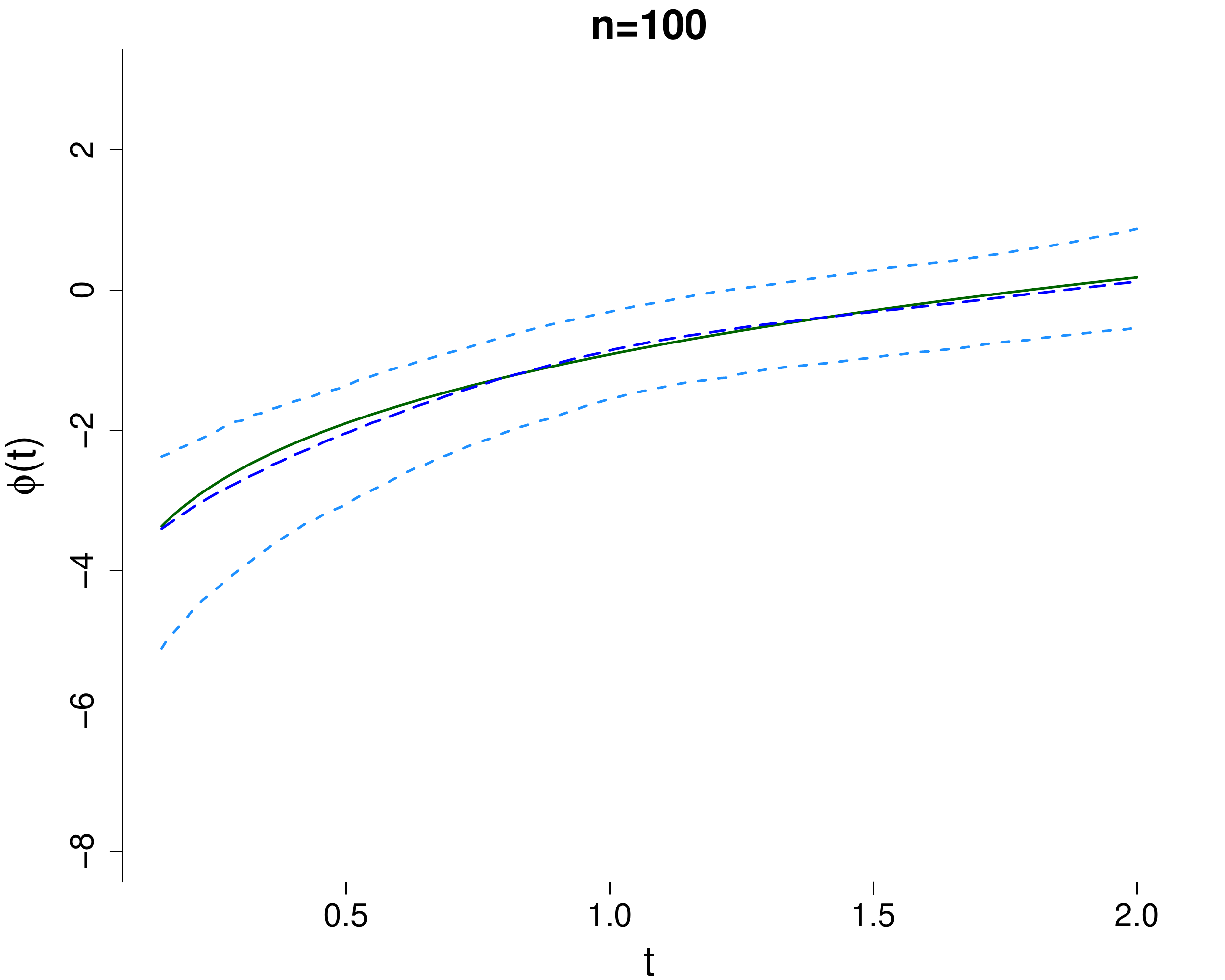}
	\includegraphics[width = 3.1 in, height = 2.8in]{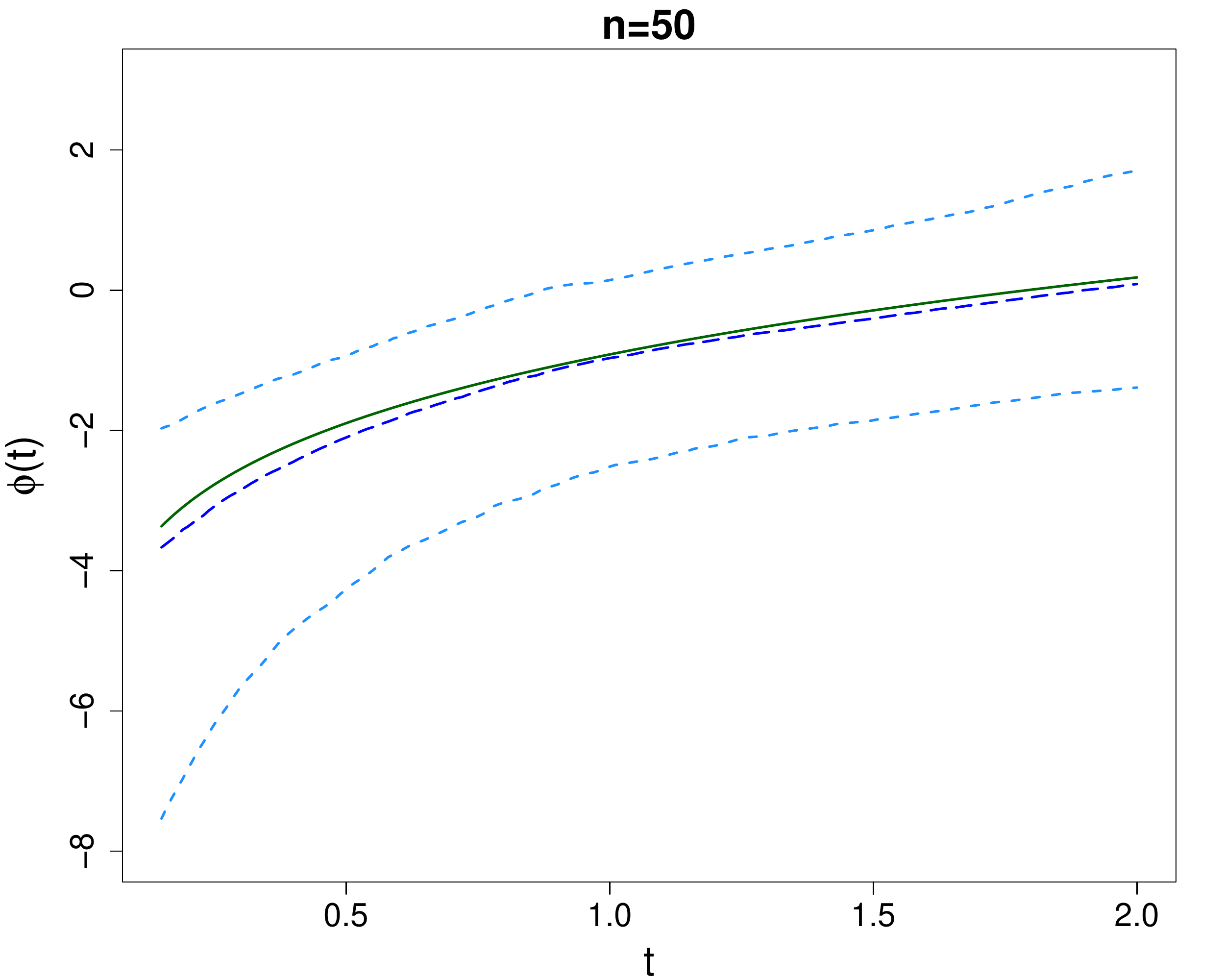}
	\includegraphics[width = 3.1 in, height = 2.8in]{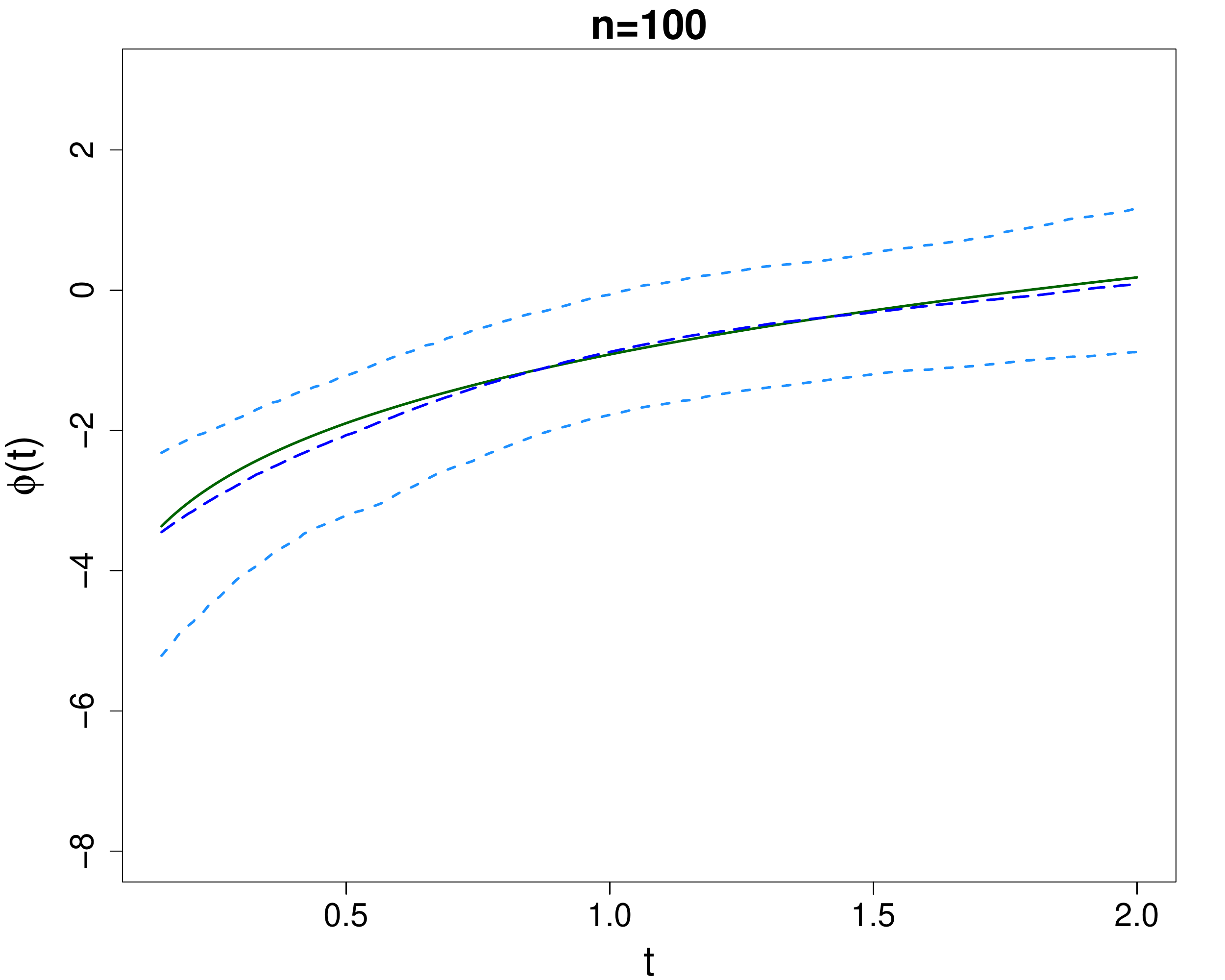}
\end{center}
{\bf Figure 2}. Summary of estimate of  $\vp(t)$  based on 1,000 replications for the PH ($\alpha = 0$) (top row) and the PO ($\alpha = 1$) models (bottom row) with $n=50$ (left) and $n=100$ (right) under C1. Pointwise means (long dashed curve), 2.5\% and 97.5\% percentiles (dashed curve) of the estimates, and the true value of the function (solid curve) are included. 


\clearpage
\noindent {\bf Table 3}.  Results of robustness study for the PH ($\alpha = 0$)  and the  PO ($\alpha = 1$) models under C1.  The PH models were employed for data generated from linear transformation models with $\alpha=0.2$, while the PO models were applied for data generated from linear transformation models with $\alpha=0.8$.  Results include empirical bias, standard deviation (SD), average of estimated standard errors (ASE), and estimated coverage probability (CP95) of 95\% confidence interval obtained from 1,000 replications with $n=50$ or 100. 
\vspace{-0cm}
\label{table2}
\vspace{-0.35cm}
\begin{center}
\begin{singlespace}
{
\begin{tabular}{ll rrrrrr}
	\hline
	\hline \\[-0.8em]
	& &\multicolumn{2}{c}{PH}& \multicolumn{2}{c}{PO}\\
	\cmidrule(r){3-4}\cmidrule(l){5-6}
	& &$\beta_1$   & $\beta_2$ & $\beta_1$ & $\beta_2$ \\[0.5em]
	$n =50$                
	& Bias  &-0.122 &-0.145    &-0.169 &-0.208\\
	& SD    &0.867 &0.542    &1.104 &0.673\\
	& ASE   &0.744 &0.440    &0.936 &0.543\\
	& CP95  &95.0\%&94.6\%  &95.2\%&95.7\%\\[0.5em]
	\hline \\ [-0.5em]
	$n =100       $   
	& Bias  &-0.035 &-0.013   &-0.111 &-0.117\\
	& SD    &0.503 &0.2952    &0.660 &0.377\\
	& ASE   &0.471 &0.267    &0.612 &0.346\\
	& CP95  &94.9\%&93.4\%  &94.9\%&94.7\%\\[0.5em]
	\hline\hline
\end{tabular}
}
\end{singlespace}
\end{center}

\clearpage
\noindent {\bf Table 4}.  Results of the breast cosmesis study under the PH ($\alpha = 0$; from both competing and penalized approach) and the PO ($\alpha = 1$; only from penalized approach) models.  
\vspace{-0.3cm}
\label{table_data1}
\vspace{-0.35cm}
\begin{center}
\begin{singlespace}
{
\begin{tabular}{llccc}
	\hline
	\hline \\[-0.8em]
	& & Competing & \multicolumn{2}{c}{Penalized}\\
	& & Approach &  \multicolumn{2}{c}{Approach}\\ \cmidrule(r){3-3} \cmidrule(l){4-5}
	& &PH & PH   & PO
	\\[0.5em]
	& Estimate & 0.896    & 0.917          & 1.042\\
	Treatment  & SE     & 0.286   & 0.285          & 0.405\\
	& 95\% CI  & $(0.335, 1.457)$ & $(0.358, 1.476)$ & $(0.249, 1.836)$\\[0.5em]
	\hline\hline
\end{tabular}
}
\end{singlespace}
\end{center}

\clearpage
\noindent {\bf Table 5} Results of signal tandmobiel study under the  PH ($\alpha = 0$; from both competing and penalized approach) and the PO ($\alpha = 1$; only from penalized approach) models.  
\vspace{-0.5cm}
\label{table_data2}
\vspace{-0.35cm}
\begin{center}
\begin{singlespace}
{
\begin{tabular}{ll llll}
	\hline
	\hline \\[-0.8em]
	& &Boy   & Attending   & Attending  & Starting age of\\
	& &   & community school               & province/council school                     & brushing the teeth 
	\\[0.5em] \hline \\ [-0.5em]
	\multicolumn{2}{l}{Competing Approach} &&&& \\[0.5em]
    PH & Estimate & -0.085 & 0.169 & 0.118 & 0.138 \\
	& SE       & 0.066  & 0.103 & 0.084 & 0.029  \\
	& 95\% CI  & $(-0.215, 0.045)$ & $(-0.033, 0.371)$& $(-0.047, 0.284)$& $(0.081, 0.196)$\\[0.5em] \hline \\ [-0.5em]
	\multicolumn{2}{l}{Penalized Approach} &&&& \\[0.5em]
    PH & Estimate & -0.085 & 0.168 & 0.118 & 0.138 \\
	& SE       & 0.066  & 0.103 & 0.084 & 0.029  \\
	& 95\% CI  & $(-0.215, 0.045)$ & $(-0.034, 0.369)$& $(-0.047, 0.283)$& $(0.081, 0.195)$\\[0.5em]
	\cline{2-6} \\ [-0.5em]
	PO & Estimate & -0.109 & 0.198 & 0.140 & 0.159 \\
	& SE       & 0.077  & 0.120 & 0.098 & 0.034  \\
	& 95\% CI  & $(-0.259, 0.041)$ & $(-0.037, 0.433)$& $(-0.052, 0.331)$& $(0.093, 0.226)$\\[0.5em]
	\hline\hline
\end{tabular}
}
\end{singlespace}
\end{center}

\newpage
\begin{center}
	\includegraphics[width = 3.2 in, height = 2.8in]{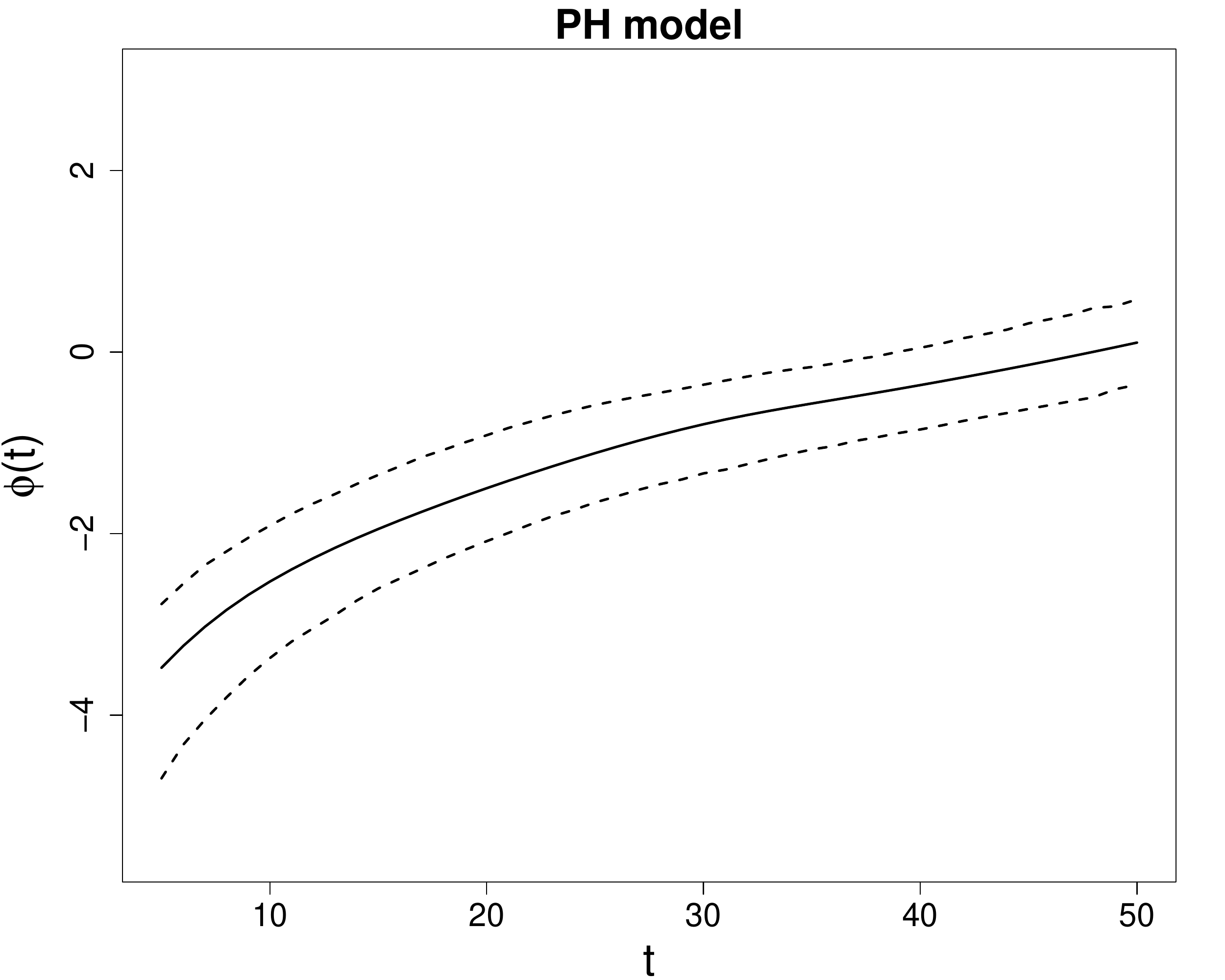}
	\includegraphics[width = 3.2 in, height = 2.8in]{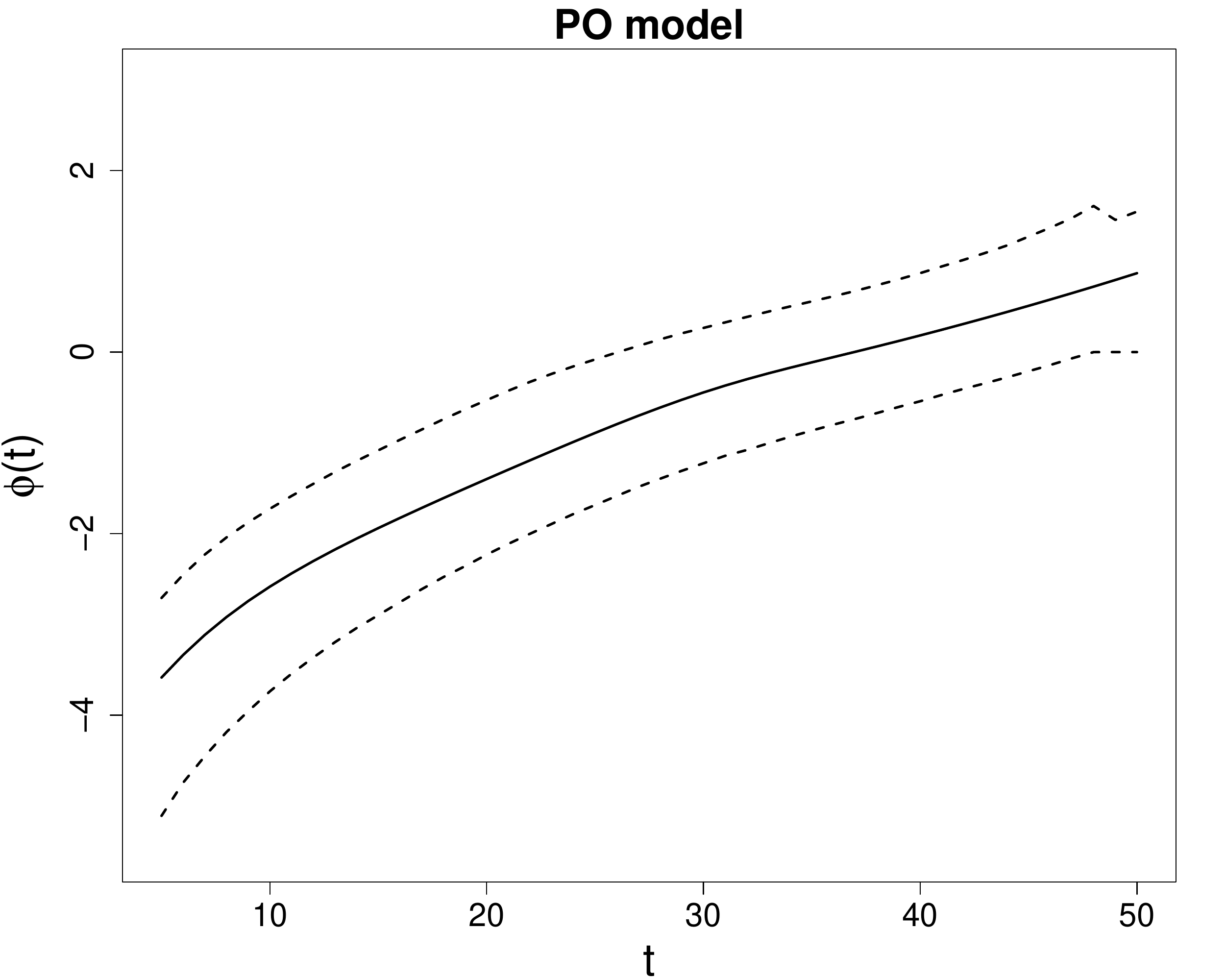}
\end{center}
{\bf Figure 3}. Estimates and corresponding 95\% pointwise confidence intervals of  transformation function  $\varphi(t)$ under the PH ($\alpha = 0$) and the PO ($\alpha = 1$)  models for breast cosmesis study. Pointwise estimates (solid curves) and 2.5\% and 97.5\% quantiles (dashed curves) based on 1,000 bootstrapped samples are included. 

\newpage
\begin{center}
	\includegraphics[width = 3.2 in, height = 2.8in]{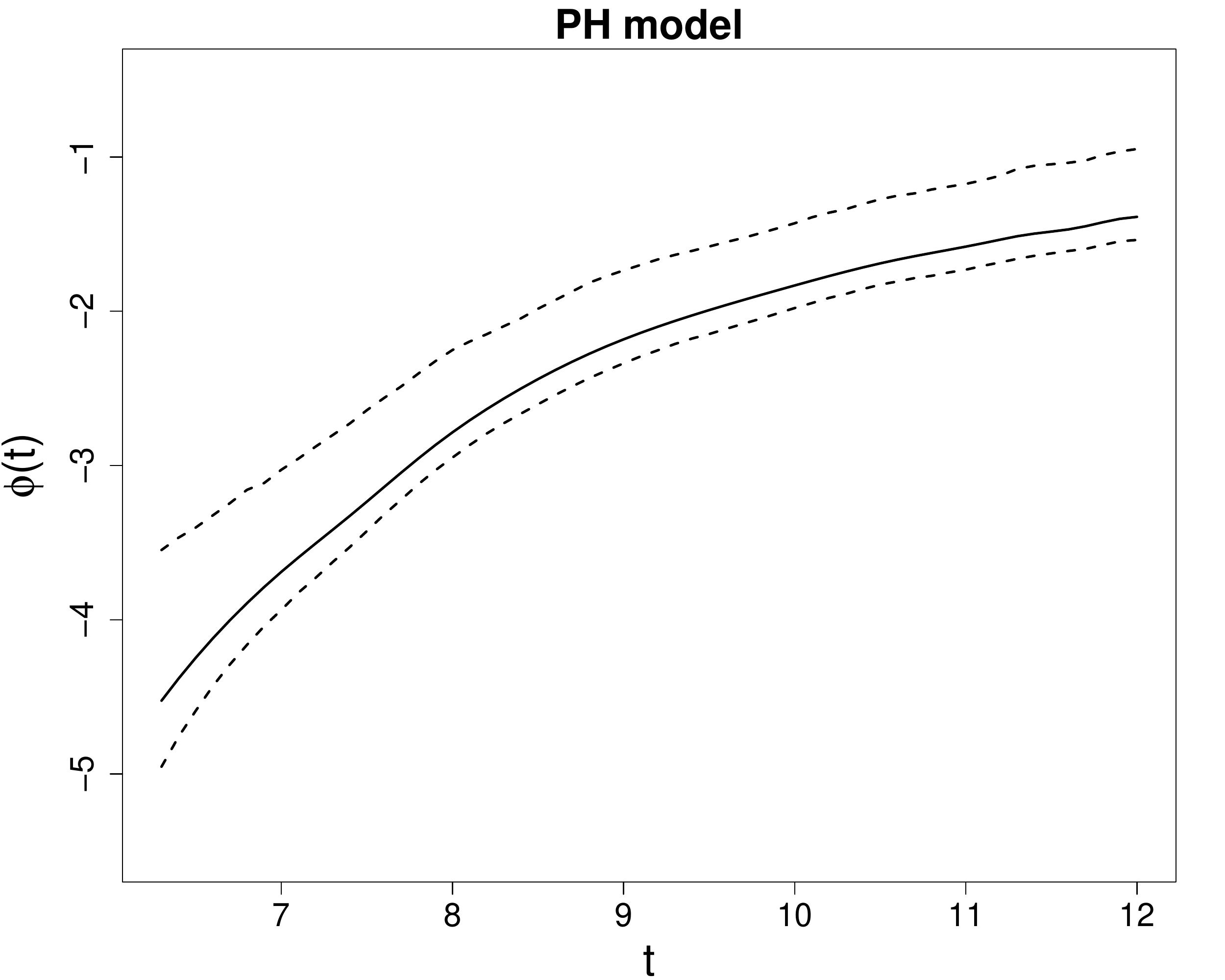}
	\includegraphics[width = 3.2 in, height = 2.8in]{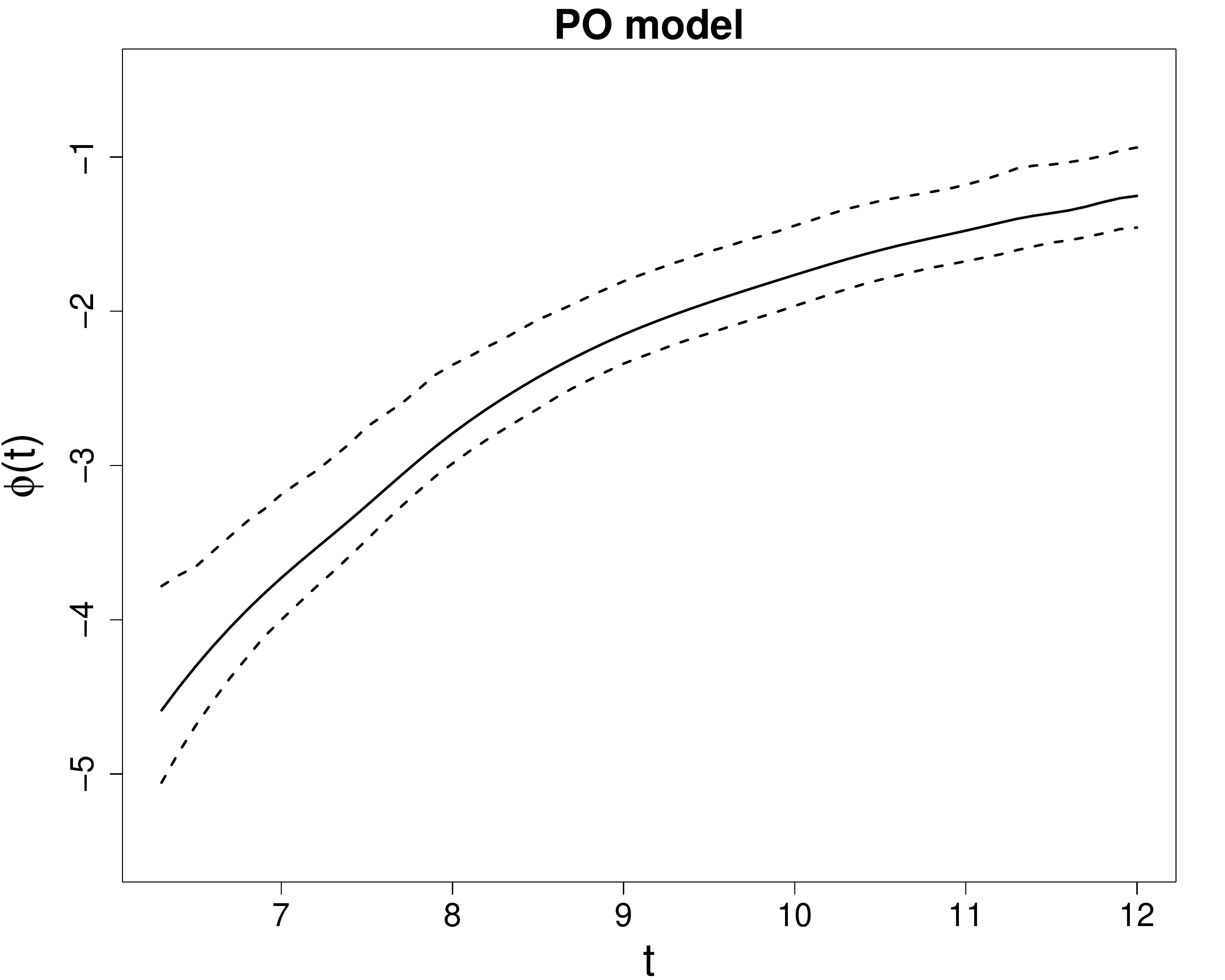}
\end{center}
{\bf Figure 4}. Estimates and corresponding 95\% pointwise confidence intervals of transformation function $\varphi(t)$ under both the PH ($\alpha = 0$) and the PO ($\alpha = 1$) models for signal tandmobiel study. Pointwise estimates (solid curves) and 2.5\% and 97.5\% quantiles (dashed curves) based on 1000 bootstrapped samples are included. 


\end{document}